\documentclass[11pt,a4paper]{article}
\usepackage{times,latexsym}
\usepackage[T1]{fontenc}
\usepackage[utf8]{inputenc}
\usepackage{url}
\usepackage{hyphenat}

\usepackage[acceptedWithA]{tacl2021v1}
\usepackage[table]{xcolor}
\usepackage{microtype}
\usepackage{inconsolata}
\usepackage{amsmath,amssymb}
\usepackage{mathtools}
\allowdisplaybreaks
\usepackage{graphicx}
\graphicspath{{./figures/}}
\usepackage{enumitem}
\usepackage{siunitx}
\usepackage[section]{placeins}
\usepackage{subcaption}
\usepackage{booktabs}
\usepackage{textgreek}
\usepackage{tabularx,array,multirow,makecell,colortbl,multicol}
\usepackage{cleveref}
\let\oldCref\Cref
\renewcommand{\Cref}[1]{\textbf{\oldCref{#1}}}
\usepackage{framed}
\usepackage{pifont}
\usepackage{tikz}
\usepackage{threeparttable}
\usepackage{rotating}
\usepackage{adjustbox}
\usepackage[most]{tcolorbox}
\usepackage{fontawesome5}

\definecolor{tblGoodLt}{HTML}{D5E8ED}   
\definecolor{tblGoodDk}{HTML}{A1CCD8}   
\definecolor{tblBadLt}{HTML}{F0D3DD}    
\definecolor{tblBadDk}{HTML}{DCA0B0}    
\definecolor{tblNeutral}{HTML}{ECEDEF}  
\definecolor{grpBand}{HTML}{C5C9CC}
\definecolor{tblBorder}{HTML}{D0D4D8}   
\definecolor{hdrBg}{HTML}{F4F6F8}
\definecolor{hdrIcon}{HTML}{4A8896}     
\definecolor{hdrSub}{HTML}{7A8A90}      
\newcommand{\rcell}[2]{%
  \tikz[baseline=(X.base)]{%
    \node[fill=#1, rounded corners=3pt,
          inner xsep=4pt, inner ysep=1.5pt,
          minimum height=3.2ex] (X) {#2};}%
}

\title{\textsc{LlaMADRS}: Evaluating Open-Source LLMs on Real Clinical Interviews---To Reason or Not to Reason?}

\author{
Gaoussou Youssouf Kebe$^{1}$ \quad Jeffrey M. Girard$^{2}$ \quad Einat Liebenthal$^{3}$, \\
\textbf{Justin Baker}$^{3}$ \quad \textbf{Fernando De~la~Torre}$^{1}$ \quad \textbf{Louis-Philippe Morency}$^{1}$ \\
$^{1}$Carnegie Mellon University, School of Computer Science, Pittsburgh, PA, USA \\
$^{2}$University of Kansas, Department of Psychology, Lawrence, KS, USA \\
$^{3}$McLean Hospital, Harvard Medical School, Boston, MA, USA \\
\texttt{\{gyk,ftorre,morency\}@cs.cmu.edu} \\
\texttt{jmgirard@ku.edu} \\
\texttt{\{eliebenthal,jtbaker\}@partners.org}
}

\begin{document}

\maketitle

\begin{abstract}
Large language models (LLMs) excel on many NLP benchmarks, but their behavior on real-world, semi-structured prediction remains underexplored. We present \textsc{LlaMADRS}, a benchmark for structured clinical assessment from dialogue built on the \textsc{CAMI} corpus of psychiatric interviews, comprising $5{,}804$ expert annotations across $541$ sessions. We evaluate $25$ open-source models (standard and reasoning-augmented; $0.6$B--$400$B parameters) and generate over $400{,}000$ predictions. Our results demonstrate that strong open-source LLMs achieve item-level accuracy with residual error below clinically substantial thresholds. Additionally, an Item-then-Sum (\textsc{ItS}) strategy, assessing symptoms individually through discrete LLM calls before synthesizing final scores, significantly reduces error relative to Direct Total Score (\textsc{DTS}) prediction across most model architectures and scales, despite reasoning models attempting similar decomposition in the reasoning traces of their \textsc{DTS} predictions. In fact, we find that performance gains attributed to ``reasoning'' depend fundamentally on prompt design: standard models equipped with structured task definitions and examples match reasoning-augmented counterparts. Among the latter, longer reasoning traces correlate with reduced error; while higher model scale does across both architectures. Our results clarify when and why reasoning helps and offer actionable guidance for deploying LLMs in semi-structured clinical assessment.
\end{abstract}

\section{Introduction}

Mental health disorders are a leading cause of disability worldwide and a major public health challenge. Even in highly developed nations, a shortage of trained clinicians leaves many individuals without timely care: more than half of the U.S. population lives in designated mental health professional shortage areas, hindering intervention and exacerbating disparities~\citep{nguyen2025counselingbench}. Depression, characterized by persistent low mood and anhedonia, remains highly prevalent. The Montgomery--\AA{}sberg Depression Rating Scale (\textsc{MADRS}) is a clinician-administered instrument comprising ten symptom domains, each rated on a $0$--$6$ scale~\citep{Montgomery1979}.

Large language models have achieved state-of-the-art performance on diverse natural language tasks, but their alignment with mental health competencies remains underexplored~\citep{na2025psychotherapy}. Emerging analyses reveal that LLMs often rely on surface-level pattern recognition rather than genuine reasoning, leading to brittleness on tasks requiring structured inference~\citep{jin2025reasoning}. Reasoning augmentation can improve performance on arithmetic or logical puzzles, but benefits are inconsistent and, in some settings, longer reasoning traces \emph{decrease} accuracy~\citep{jin2025reasoning}. In mental health domains, reasoning-driven prompting has been proposed to enhance classification accuracy~\citep{teng2025cot}, yet improvements are modest and dataset-specific: chain-of-thought and related strategies yield notable gains on some tasks while failing to generalize to others~\citep{patil2025cognitive}. These observations motivate our systematic comparison of reasoning-augmented and standard LLMs on clinical interviews.

Our contributions advance both computational psychiatry and natural language processing:

\begin{figure*}
    
    \centering
  \includegraphics[width=\linewidth]{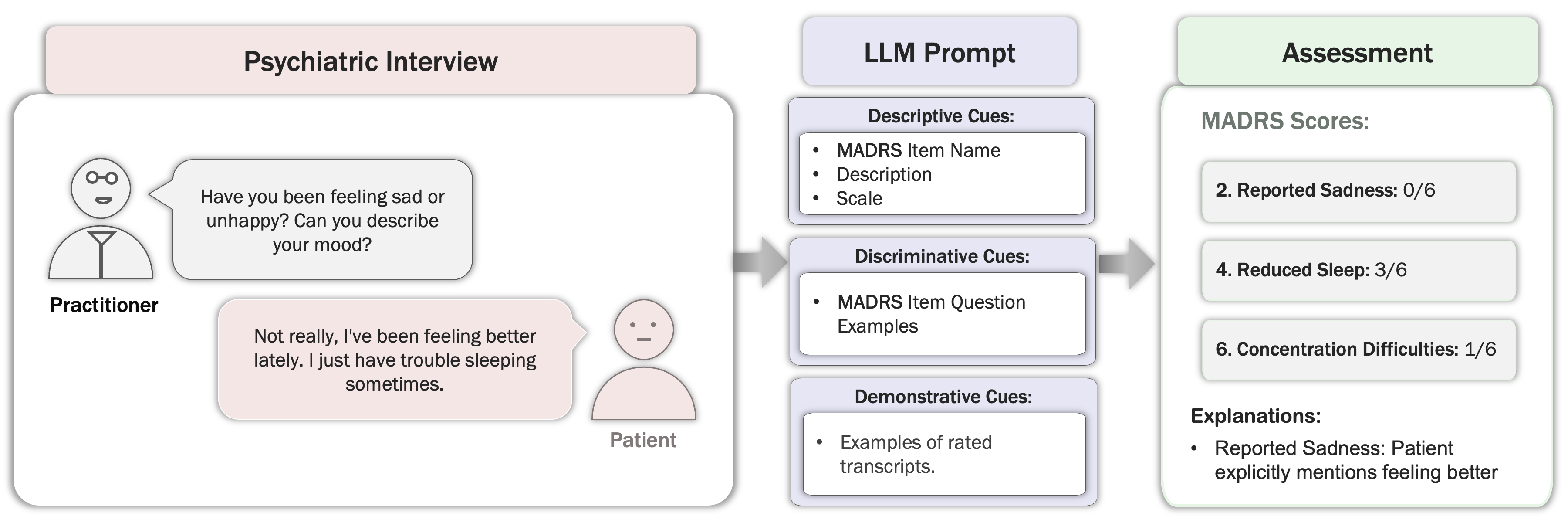}

\caption{\textbf{Overview of the \textsc{LlaMADRS} framework.} Left: a structured clinical interview between a patient and clinician. Right: automated depression assessment using a large language model, including scoring of \textsc{MADRS} items with item-wise explanations.}
\label{fig:teaser}
\end{figure*}

\begin{enumerate}[leftmargin=*,nosep]
\item \textbf{\textsc{LlaMADRS}}, a benchmark of $5{,}804$ \textsc{MADRS} assessments across $541$ real patient--clinician interview sessions from the \textsc{CAMI} corpus.
\item \textbf{Clinically reliable accuracy}: Strong LLMs achieve item-level Mean Absolute Error (MAE) in the moderate range ($0.6$--$1.2$), with total-scale Mean Absolute Error (MAE) in the clinically acceptable range ($< 6$) under Item-then-Sum (\textsc{ItS}) scoring.
\item \textbf{When reasoning helps}: Explicit reasoning helps when scaffolding is sparse, but provides limited benefit under well-structured prompts with clinical descriptive cues, where standard models frequently match or surpass reasoning-augmented variants.
\item \textbf{Key performance predictors}: Model scale reliably improves accuracy across both reasoning and non-reasoning architectures, whereas longer reasoning traces appear to benefit reasoning-augmented models.
\end{enumerate}
\section{Related Work}
\subsection{NLP and Mental Health}
Early applications of NLP for mental health focused on detecting depression, anxiety, and suicidality in social media data using bag-of-words, topic models, and early transformers~\citep{coppersmith2014quantifying,dechoudhury2013predicting,eichstaedt2018facebook,shen2017detecting,ji-etal-2022-mentalbert}. With LLMs, research broadened from binary screening to multi-label symptom detection, information extraction, and explanation generation across single posts, user histories, and dialogues~\citep{yang2023interpretable,yang2024mentallama,xu2024mental,bao2024explainable,raihan2024mentalhelp,mohammadi2024welldunn,schirmer2024ptsd,skianis2024severity}. Clinical-facing efforts include multi-turn interview analysis for PTSD and broader symptom delineation/summarization~\citep{tu2024ptsd,so2024psychiatric}, alongside severity estimation via taxonomy-aligned summaries (e.g., BDI)~\citep{aragon2024bdi,wang2024e-bdi}. NLP-based evaluations of clinician competencies have also tried to detect gaps in empathy, cultural sensitivity, and ethics, with results underscoring the need for clinically grounded tasks and item-level analyses rather than aggregate labels~\citep{nguyen2025counselingbench}.
\begin{table*}
\centering
\footnotesize
\caption{Comparison of LLM-based psychiatric assessment works.}
\label{tab:related_work}
\renewcommand{\arraystretch}{1.02}
\setlength{\tabcolsep}{6pt}
\begin{adjustbox}{max width=\textwidth}
\begin{tabular}{llll}
\toprule
\textbf{Study} & \textbf{Models} & \textbf{Dataset / $N$} & \textbf{Population} \\
\midrule
\rowcolor{blue!8}\textbf{This work} & \textbf{$25$ open-source (R \& NR)} & \textbf{\textsc{CAMI} / $541$} & \textbf{Clinician interviews; inpatients} \\
\midrule
\citet{tu2024ptsd} & GPT-4, Llama-2 & Clinical interviews / $411$ & Trauma interviews \\
\citet{so2024psychiatric} & GPT-4 Turbo; GPT-3.5 (FT) & Psychiatric interviews (KR) & Clinical \\
\citet{galatzerlevy2023llmpsy} & Med-PaLM~2 (ZS) & Clinical descriptions & Screening \\
\citet{arcan2024assessment} & ChatGPT, Llama-2 (ZS) & DAIC-WOZ, etc. & Community/elicited \\
\citet{aragon2024bdi} & GPT-3.5/4 (ZS) & Social media (per-user) & General population \\
\citet{yang2023interpretable} & GPT-3.5/LLaMA & $11$ social-media sets & General population \\
\citet{xu2024mental} & Inst.-tuned LLM (FT) & Online text (various) & General population \\
\citet{skianis2024severity} & LLMs (ZS) & $6$ languages (new) & Cross-lingual social media \\
\bottomrule
\end{tabular}
\end{adjustbox}
\end{table*}

\subsection{Reasoning-Augmented LLMs in Mental Health}
Debates persist on whether LLMs \emph{reason} or perform sophisticated pattern matching. Evidence shows that explicit chain-of-thought (CoT), self-consistency, and related strategies yield mixed and dataset-specific gains; longer reasoning traces may even degrade accuracy~\citep{jin2025reasoning,teng2025cot,patil2025cognitive}. In psychotherapy and mental-health NLP, prompting styles (emotion prompting, CoT, role prompting, multi-agent debate) help interpretability or specific subtasks but do not guarantee robust generalization across settings or constructs~\citep{yang2023interpretable,lim2024erd,singh2024suicide,uluslu2024suicidality}. These results motivate head-to-head evaluation of reasoning-augmented vs. standard open LLMs under clinically moderate objectives.

\subsection{LLMs for Clinical Assessment and Severity Scoring}
A stream of work examines LLMs for clinical assessment and severity estimation from interviews or consolidated text. Med-PaLM~2 showed zero-shot alignment with clinician ratings for depression severity (and limited generalization to PTSD)~\citep{galatzerlevy2023llmpsy}. On DAIC-WOZ and related corpora, general LLMs exhibit moderate text-regression performance relative to specialized transformers~\citep{arcan2024assessment}.

Compared to prior work that focuses on social media, synthetic datasets, or aggregate screening~\citep{tu2024ptsd,so2024psychiatric,galatzerlevy2023llmpsy,arcan2024assessment,yang2023interpretable,xu2024mental,aragon2024bdi,skianis2024severity}, we offer (1) \emph{item-level and total} \textsc{MADRS} scoring over \emph{real} clinician--patient interviews; (2) a broad open-source model comparison spanning architectures, scales, and reasoning styles; and (3) clinically grounded error analyses and prompt ablations disentangling descriptive cues, demonstrative cues, and reasoning augmentation.

\section{Dataset: \textsc{CAMI}}
\label{sec:cami}
The Context-Adaptive Multimodal Informatics (\textsc{CAMI}) dataset \citep{Culhane2023} comprises authentic patient--clinician dialogues from inpatient psychiatric care settings. All participants provided informed consent under IRB-approved protocols. Each semi-structured session lasted approximately thirty minutes during which trained raters administered the \textsc{MADRS}. The dataset contains $541$ interviews from $277$ adult patients, totaling $5{,}804$ item-level and total-score \textsc{MADRS} ratings.

\paragraph{MADRS annotations.}
Each session was rated by a single clinically trained research assistant on the ten \textsc{MADRS} items: (I1)~Apparent Sadness, (I2)~Reported Sadness, (I3)~Inner Tension, (I4)~Reduced Sleep, (I5)~Reduced Appetite, (I6)~Concentration Difficulties, (I7)~Lassitude, (I8)~Inability to Feel, (I9)~Pessimistic Thoughts, and (I10)~Suicidal Thoughts. These clinical research assistants were trained to administer and score the \textsc{MADRS} through structured supervision, calibration, and review of recorded semi-structured interviews using standardized rating guidelines. Before independent rating, all raters were required to achieve item-level Cohen's $\kappa \geq 0.80$ on separate calibration data~\citep{Gillis2023}. Each item is scored on a seven-point scale from $0$ (absent) to $6$ (severe), where even anchor points carry specific behavioral descriptions: $0$~=~no symptoms, $2$~=~mild, $4$~=~moderate, $6$~=~severe. Odd scores ($1$, $3$, $5$) denote intermediate severity levels between adjacent anchors (e.g., $1$~=~between absent and mild; $3$~=~between mild and moderate; $5$~=~between moderate and severe). The ten item scores are summed to produce a total score ranging from $0$ to $60$, interpreted as: absent ($0$--$6$), mild ($7$--$19$), moderate ($20$--$34$), and severe depression ($35$--$60$)~\citep{Montgomery1979}. \Cref{fig:madrs-overview} shows the score distribution across all ten items alongside the total-score severity bands.

\paragraph{Audio processing.}
Transcriptions were generated using both Whisper \citep{Radford2023} and Parakeet \citep{sekoyan2025canary} ASR systems, then processed with Qwen 3-32B \citep{Wang2024diarization} for diarization to assign utterances to speakers. For item-level prediction, Qwen 3-32B was used to segment interviews by identifying question--response exchanges corresponding to specific \textsc{MADRS} items.

\paragraph{Availability.}
All experiments were conducted on the original, non-de-identified transcripts under the IRB-approved protocol governing the \textsc{CAMI} dataset. For external access, de-identified versions of the transcripts, clinician ratings, and session-level metadata will be available under controlled access.\footnote{For data access requests, please contact Justin Baker, McLean Hospital, Harvard Medical School, Boston, MA, USA, at \texttt{jtbaker@mgb.org}. Code is available at \url{https://github.com/llamadrs/llamadrs}.} Access is limited to credentialed researchers who agree to a Data Use Agreement.

\section{Approach}
We evaluate $25$ open-source LLMs spanning eight architectural families and parameter counts from $0.6$ billion to $400$ billion (dense or mixture-of-experts variants).

\subsection{Prompting Framework}
\label{ssec:prompt-framework}
Each prompt comprises three components: (1) a task description with clinical definitions and severity anchors (descriptive cues); (2) annotated interview excerpts providing rating rationale examples (demonstrative cues); and (3) JSON schema-enforced output requirements. Prompt templates are provided in \textbf{Appendix} \Cref{sec:appendix-prompts}.

For item-level predictions, we provide the relevant \textsc{MADRS} item definition and severity anchors (as described in \Cref{sec:cami}), along with the corresponding transcript segment for that item.

For total-score prediction, we provide an overview of all ten \textsc{MADRS} items and refer to the standard total-score interpretation, also defined in \Cref{sec:cami}, then ask the model to output a single total score. The complete \textsc{MADRS} interview transcript is used for total score prediction.

\begin{figure*}[t]
  \centering
  \includegraphics[width=\linewidth]{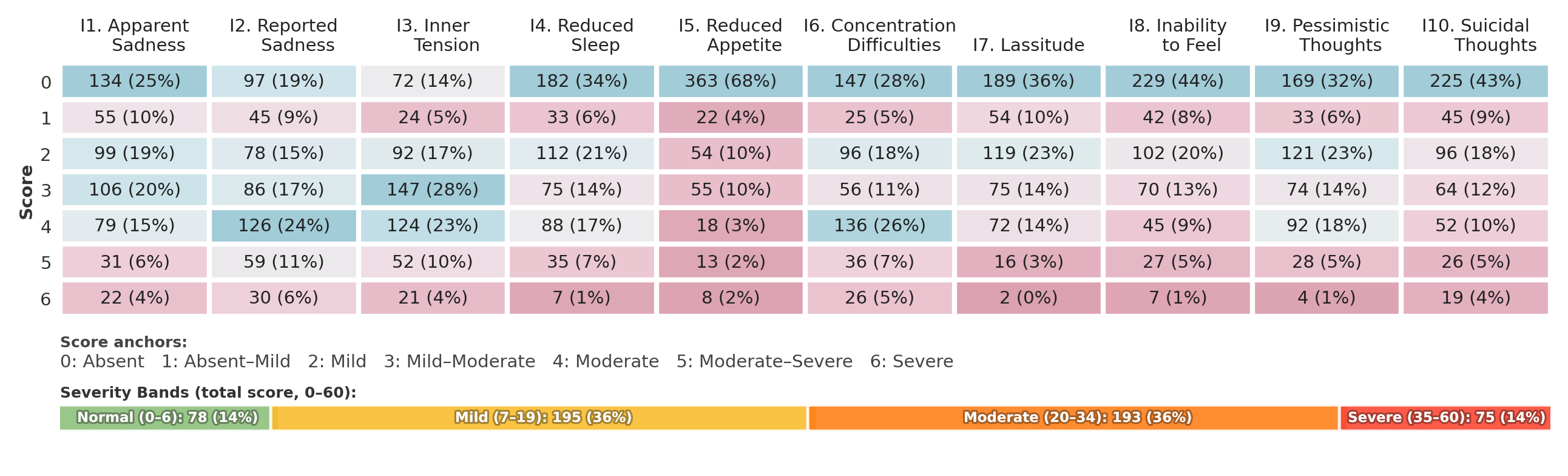}
  \caption{\textbf{\textsc{MADRS} score distributions.} Per-item score counts (with percentages) across the $541$ sessions and overall severity-band distribution (No depression $0$--$6$, Mild $7$--$19$, Moderate $20$--$34$, Severe $35$--$60$).}
  \label{fig:madrs-overview}
\end{figure*}

\subsection{Evaluation Protocol}
\paragraph{Error metric.} We adopt Mean Absolute Error (MAE) as our primary evaluation metric, defined as $\mathrm{MAE} = \frac{1}{N}\sum_{i=1}^{N}|\hat{y}_i - y_i|$, where $\hat{y}_i$ and $y_i$ are predicted and gold-standard ratings, respectively. MAE is preferred over squared-error metrics because it is directly interpretable in the original rating units (i.e., a MAE of $1.0$ corresponds to an average misprediction of one scale point) and aligns naturally with clinical minimal-change thresholds~\citep{Turkoz2021}.

\paragraph{Seed replication and averaging.} For each model--session pair, we generate $3$ independent predictions using different random seeds. MAE is computed per seed, then averaged across the three seeds to produce the reported seed-averaged MAE. Cross-seed standard deviations are shown as $\pm$ values throughout Tables~\ref{tab:table2_total_scores}--\ref{tab:table3_item_scores}.

\paragraph{Sum-based aggregation.} We evaluate two strategies for obtaining total \textsc{MADRS} scores. \textbf{Direct total score (\textsc{DTS})}: the model receives the full interview transcript and predicts the total score ($0$--$60$) in a single inference call. \textbf{Item-then-sum (\textsc{ItS})}: the model scores each of the ten items independently via separate inference calls, and the total is computed by mechanically summing the ten item-level predictions, i.e., $\hat{T}_{\mathrm{ItS}} = \sum_{i=1}^{10} \hat{y}_i$, with no model involvement in the summation step.

\paragraph{Clinical error thresholds.} We adopt thresholds from \citet{Turkoz2021} to interpret prediction errors: a less than 6-point change on the MADRS total scale (0--60) is ``clinically acceptable,'' while more than 12 points is ``clinically substantial.'' We repurpose these as error evaluation anchors, defining three bands: \textbf{clinically acceptable} (MAE $<$ 6 total, $<$ 0.6 per item), \textbf{moderate error} (MAE 6--12 total, 0.6--1.2 per item), and \textbf{substantial error} (MAE $\geq$ 12 total, $\geq$ 1.2 per item). Errors below 6 points are unlikely to affect clinical interpretation, while errors exceeding 12 points may alter treatment decisions.

\paragraph{Handling invalid outputs.}
Predictions may fail due to (a)~unparseable JSON output (e.g., malformed syntax), (b)~absent or out-of-range rating fields (e.g., returning a rating of $7$ or omitting the score entirely), or (c)~generation failures (e.g., incomplete outputs due to context-length truncation). For such cases, we apply a conservative backoff: we assign the maximum absolute error ($6$ item-level; $60$ total), so models are not rewarded for invalid or selectively abstained outputs.

\paragraph{Screening evaluation.}
For screening, we binarize item scores at $\geq 3$ and total scores at $\geq 20$, reporting F$_1$ as the primary metric. The total-score cutoff of $\geq 20$ corresponds to the standard \textsc{MADRS} mild--moderate boundary~\citep{Montgomery1979} and produces a near-balanced split in \textsc{CAMI} (see \Cref{fig:madrs-overview}): $273/541$ ($50.5\%$) $< 20$ vs.\ $268/541$ ($49.5\%$) $\geq 20$.

\subsection{Prompt Ablation}
\label{ssec:ablation-methods}
Building on the prompting framework in \Cref{ssec:prompt-framework}—(1) clinical task and severity descriptive cues, (2) annotated demonstrative cues, and (3) schema-enforced outputs—we ablate the first two components to assess whether descriptive cues and demonstrative cues differentially benefit LLMs. In parallel, we run a paired, item-level comparison of \textsc{MADRS} performance between the reasoning-augmented and standard variants of \textbf{Qwen~3~Next (80B; MoE)} across four configurations: \textbf{All} (descriptive cues + demonstrative cues), \textbf{No~Descriptions} (demonstrative cues only), \textbf{No~Demonstrations} (descriptive cues only), and \textbf{Raw} (no cues).

\subsection{Statistical Analysis of Prediction Errors}
\label{ssec:stats}

We fit two cross-classified linear mixed-effects models~\citep{Browne2001,Goldstein2010} for standard and reasoning-augmented LLMs.

\textbf{Models.} For non-reasoning models:
\begin{equation}
\label{eq:lmm_nonreason}
\begin{aligned}
\mathrm{Y}_{p,s,m,\tau}^{(\text{NR})} ={}& \alpha
+ \beta_{1} S^{W}_{p,s,\tau}
+ \beta_{2} S^{B}_{p,\tau}
+ \beta_{3} T^{W}_{p,s}
+ \beta_{4} T^{B}_{p} \\
&{}\; + \beta_{5}\,\mathrm{Par}_{m}
+ \beta_{6}\,\mathrm{Cont}_{m}
+ \beta_{7}\,\mathrm{MOE}_{m} \\
&{}\; + b_{m} + u_{p} + v_{s:p} + w_{\tau} + \varepsilon_{p,s,m,\tau}.
\end{aligned}
\end{equation}

For reasoning models, we add reasoning-token predictors:
\begin{equation}
\label{eq:lmm_reason}
\mathrm{Y}_{p,s,m,\tau}^{(\text{R})}
= \mathrm{Y}_{p,s,m,\tau}^{(\text{NR})}
+ \beta_{8} R^{W}_{p,s,\tau}
+ \beta_{9} R^{B}_{p,\tau}.
\end{equation}

\textbf{Variables.} $\mathrm{Y}_{p,s,m,\tau}$ is seed-averaged MAE for patient $p$, session $s$, model $m$, task $\tau$ (ten \textsc{MADRS} items plus total, with total-score errors rescaled to item range by dividing by $10$). Continuous predictors are log-transformed, $z$-standardized, and decomposed into within- and between-patient components: $S_{p,s,\tau}$ (session--item severity), $T_{p,s}$ (transcript tokens), $R_{p,s,\tau}$ (reasoning tokens; $R=0$ for non-reasoning). Model covariates are $\log_{10}$ parameters ($\mathrm{Par}_{m}$), $\log_{10}$ context window ($\mathrm{Cont}_{m}$), and mixture-of-experts indicator ($\mathrm{MOE}_{m}$). Random intercepts $b_m$, $u_p$, $v_{s:p}$, $w_\tau$ account for model, patient, session-within-patient, and item variance; $\varepsilon_{p,s,m,\tau}$ are homoscedastic Gaussian residuals.
\section{Results}

Our evaluation addresses three central questions: (1) Can LLMs score \textsc{MADRS} from real interviews with clinically reliable accuracy? (2) Under what conditions does reasoning improve performance over standard models? (3) What factors predict model performance on clinical assessment tasks?

\begin{table*}[!tb]
\centering
\footnotesize
\caption{Total-score evaluation: reasoning vs.\ non-reasoning. \textbf{DTS}: $\hat{T}{=}f_{\theta}(x)$; \textbf{ItS}: $\hat{T}{=}\sum\hat{y}_i$. MoE sizes: Active--Total params. \textbf{Bold}\,=\,best; darker\,=\,best/worst. \textbf{MAE}: \protect\colorbox{tblGoodLt}{\strut Acceptable} (${<}$\,6), \protect\colorbox{tblBadLt}{\strut Substantial} (${\geq}$\,12). \textbf{F1} ($T{\geq}20$): \protect\colorbox{tblGoodLt}{\strut ${\geq}$\,Q3} (0.86), \protect\colorbox{tblBadLt}{\strut ${<}$\,Q1} (0.77).}
\label{tab:table2_total_scores}
\begin{tcolorbox}[
  enhanced,
  boxrule=0.5pt,
  colframe=tblBorder,
  colback=white,
  arc=8pt,
  outer arc=8pt,
  left=3pt, right=3pt, top=3pt, bottom=3pt,
  boxsep=0pt,
  before upper={\arrayrulecolor{tblBorder}\renewcommand{\arraystretch}{1.35}\setlength{\tabcolsep}{3pt}},
]
\begin{tabularx}{\linewidth}{@{} >{\hsize=0.46\hsize\raggedright\arraybackslash}X >{\hsize=0.12\hsize\centering\arraybackslash}X >{\hsize=0.12\hsize\centering\arraybackslash}X >{\hsize=0.17\hsize\centering\arraybackslash}X >{\hsize=0.17\hsize\centering\arraybackslash}X >{\hsize=0.17\hsize\centering\arraybackslash}X >{\hsize=0.17\hsize\centering\arraybackslash}X @{}}
\arrayrulecolor{tblBorder}
\rcell{tblNeutral}{\textsf{Model}\,{\scriptsize\textcolor{hdrSub}{(Size)}}} & \rcell{tblNeutral}{\textsf{Arch.}} & \rcell{tblNeutral}{\textsf{Ctx.}} & \multicolumn{2}{c}{\rcell{tblNeutral}{\textsf{MAE}\,{\scriptsize$\downarrow$}}} & \multicolumn{2}{c}{\rcell{tblNeutral}{\textsf{F1}\,{\scriptsize$\uparrow$}}} \\
\addlinespace[3pt]
\rowcolor{tblNeutral} \multicolumn{3}{@{}l@{}}{\small\textsc{Reasoning Models}} & \textsc{DTS} &  \textsc{ItS} & \textsc{DTS} &  \textsc{ItS}\\
\addlinespace[1.5pt]
Qwen 3 Next {\tiny (3B-80B)} & MoE & 262k & \rcell{tblGoodLt}{5.90 {\scriptsize $\pm$0.17}} & \rcell{tblGoodLt}{4.31 {\scriptsize $\pm$0.11}} & \rcell{tblGoodDk}{\textbf{0.87} {\scriptsize $\pm$0.01}} & \rcell{tblGoodLt}{0.87 {\scriptsize $\pm$0.01}} \\
GPT OSS 120B {\tiny (5B-117B)} & MoE & 131k & \rcell{tblNeutral}{6.80 {\scriptsize $\pm$0.19}} & \rcell{tblGoodLt}{3.78 {\scriptsize $\pm$0.08}} & \rcell{tblGoodLt}{0.84 {\scriptsize $\pm$0.00}} & \rcell{tblNeutral}{0.86 {\scriptsize $\pm$0.01}} \\
Qwen 3 {\tiny (22B-235B)} & MoE & 262k & \rcell{tblNeutral}{7.60 {\scriptsize $\pm$0.44}} & \rcell{tblGoodLt}{3.68 {\scriptsize $\pm$0.02}} & \rcell{tblGoodLt}{0.85 {\scriptsize $\pm$0.01}} & \rcell{tblNeutral}{0.87 {\scriptsize $\pm$0.00}} \\
GPT OSS 20B {\tiny (3B-21B)} & MoE & 131k & \rcell{tblNeutral}{7.86 {\scriptsize $\pm$0.18}} & \rcell{tblGoodLt}{3.99 {\scriptsize $\pm$0.03}} & \rcell{tblNeutral}{0.82 {\scriptsize $\pm$0.01}} & \rcell{tblNeutral}{0.85 {\scriptsize $\pm$0.02}} \\
Magistral Small 2507 {\tiny (24B)} & Dense & 40k & \rcell{tblNeutral}{8.76 {\scriptsize $\pm$0.90}} & \rcell{tblGoodLt}{3.81 {\scriptsize $\pm$0.02}} & \rcell{tblNeutral}{0.82 {\scriptsize $\pm$0.01}} & \rcell{tblNeutral}{0.86 {\scriptsize $\pm$0.01}} \\
DeepSeek R1 Qwen 2.5 {\tiny (32B)} & Dense & 131k & \rcell{tblNeutral}{10.03 {\scriptsize $\pm$0.78}} & \rcell{tblGoodLt}{3.83 {\scriptsize $\pm$0.04}} & \rcell{tblNeutral}{0.77 {\scriptsize $\pm$0.01}} & \rcell{tblNeutral}{0.86 {\scriptsize $\pm$0.01}} \\
DeepSeek R1 Llama 3.3 {\tiny (70B)} & Dense & 131k & \rcell{tblNeutral}{10.14 {\scriptsize $\pm$0.79}} & \rcell{tblGoodLt}{3.90 {\scriptsize $\pm$0.04}} & \rcell{tblNeutral}{0.81 {\scriptsize $\pm$0.01}} & \rcell{tblGoodLt}{0.87 {\scriptsize $\pm$0.01}} \\
Qwen 3 {\tiny (14B)} & Dense & 131k & \rcell{tblNeutral}{10.84 {\scriptsize $\pm$3.01}} & \rcell{tblGoodLt}{4.12 {\scriptsize $\pm$0.04}} & \rcell{tblNeutral}{0.79 {\scriptsize $\pm$0.05}} & \rcell{tblGoodLt}{0.88 {\scriptsize $\pm$0.01}} \\
Qwen 3 {\tiny (32B)} & Dense & 131k & \rcell{tblBadLt}{12.82 {\scriptsize $\pm$0.58}} & \rcell{tblGoodLt}{4.06 {\scriptsize $\pm$0.02}} & \rcell{tblNeutral}{0.75 {\scriptsize $\pm$0.00}} & \rcell{tblNeutral}{0.87 {\scriptsize $\pm$0.01}} \\
Qwen 3 {\tiny (8B)} & Dense & 131k & \rcell{tblBadLt}{13.56 {\scriptsize $\pm$0.60}} & \rcell{tblGoodLt}{4.43 {\scriptsize $\pm$0.05}} & \rcell{tblBadLt}{0.73 {\scriptsize $\pm$0.03}} & \rcell{tblNeutral}{0.86 {\scriptsize $\pm$0.00}} \\
DeepSeek R1 Llama 3.1 {\tiny (8B)} & Dense & 131k & \rcell{tblBadLt}{13.78 {\scriptsize $\pm$0.21}} & \rcell{tblNeutral}{11.84 {\scriptsize $\pm$0.82}} & \rcell{tblBadLt}{0.73 {\scriptsize $\pm$0.01}} & \rcell{tblBadDk}{0.71 {\scriptsize $\pm$0.01}} \\
Qwen 3 {\tiny (4B)} & Dense & 131k & \rcell{tblBadLt}{15.23 {\scriptsize $\pm$0.00}} & \rcell{tblGoodLt}{5.46 {\scriptsize $\pm$0.11}} & \rcell{tblBadLt}{0.72 {\scriptsize $\pm$0.03}} & \rcell{tblBadLt}{0.84 {\scriptsize $\pm$0.01}} \\
Qwen 3 {\tiny (1.7B)} & Dense & 40k & \rcell{tblBadLt}{19.05 {\scriptsize $\pm$0.29}} & \rcell{tblNeutral}{7.45 {\scriptsize $\pm$0.17}} & \rcell{tblBadDk}{0.57 {\scriptsize $\pm$0.02}} & \rcell{tblBadLt}{0.78 {\scriptsize $\pm$0.01}} \\
Qwen 3 {\tiny (0.6B)} & Dense & 40k & \rcell{tblBadLt}{19.72 {\scriptsize $\pm$0.91}} & \rcell{tblNeutral}{9.46 {\scriptsize $\pm$0.19}} & \rcell{tblBadLt}{0.58 {\scriptsize $\pm$0.01}} & \rcell{tblBadLt}{0.72 {\scriptsize $\pm$0.01}} \\
QwQ {\tiny (32B)} & Dense & 131k & \rcell{tblBadLt}{21.09 {\scriptsize $\pm$1.06}} & \rcell{tblGoodLt}{4.03 {\scriptsize $\pm$0.05}} & \rcell{tblNeutral}{0.76 {\scriptsize $\pm$0.01}} & \rcell{tblNeutral}{0.87 {\scriptsize $\pm$0.00}} \\
Qwen 3 {\tiny (3B-30B)} & MoE & 262k & \rcell{tblBadDk}{50.22 {\scriptsize $\pm$2.22}} & \rcell{tblGoodLt}{3.98 {\scriptsize $\pm$0.11}} & \rcell{tblBadLt}{0.67 {\scriptsize $\pm$0.05}} & \rcell{tblNeutral}{0.87 {\scriptsize $\pm$0.01}} \\
\addlinespace[1.5pt]
\rowcolor{tblNeutral} \multicolumn{3}{@{}l@{}}{\small\textsc{Non-Reasoning Models}} & \textsc{DTS} &  \textsc{ItS} & \textsc{DTS} &  \textsc{ItS}\\
\addlinespace[1.5pt]
Qwen 2.5 {\tiny (72B)} & Dense & 131k & \rcell{tblGoodDk}{\textbf{5.53} {\scriptsize $\pm$0.15}} & \rcell{tblGoodLt}{3.80 {\scriptsize $\pm$0.03}} & \rcell{tblGoodLt}{0.86 {\scriptsize $\pm$0.01}} & \rcell{tblGoodLt}{0.87 {\scriptsize $\pm$0.00}} \\
Llama 4 Maverick {\tiny (17B-400B)} & MoE & 1m & \rcell{tblGoodLt}{5.60 {\scriptsize $\pm$0.05}} & \rcell{tblGoodLt}{4.00 {\scriptsize $\pm$0.02}} & \rcell{tblNeutral}{0.84 {\scriptsize $\pm$0.00}} & \rcell{tblGoodLt}{0.88 {\scriptsize $\pm$0.00}} \\
Llama 4 Scout {\tiny (17B-109B)} & MoE & 10m & \rcell{tblGoodLt}{5.75 {\scriptsize $\pm$0.06}} & \rcell{tblGoodLt}{3.70 {\scriptsize $\pm$0.00}} & \rcell{tblNeutral}{0.82 {\scriptsize $\pm$0.02}} & \rcell{tblNeutral}{0.86 {\scriptsize $\pm$0.00}} \\
Qwen 2.5 {\tiny (14B): 1M} & Dense & 1m & \rcell{tblNeutral}{6.09 {\scriptsize $\pm$0.20}} & \rcell{tblGoodDk}{\textbf{3.61} {\scriptsize $\pm$0.03}} & \rcell{tblNeutral}{0.82 {\scriptsize $\pm$0.01}} & \rcell{tblGoodLt}{0.88 {\scriptsize $\pm$0.01}} \\
Qwen 2.5 {\tiny (7B): 1M} & Dense & 1m & \rcell{tblNeutral}{6.11 {\scriptsize $\pm$0.35}} & \rcell{tblGoodLt}{4.71 {\scriptsize $\pm$0.27}} & \rcell{tblGoodLt}{0.85 {\scriptsize $\pm$0.01}} & \rcell{tblNeutral}{0.86 {\scriptsize $\pm$0.00}} \\
Llama 3.3 {\tiny (70B)} & Dense & 131k & \rcell{tblNeutral}{7.32 {\scriptsize $\pm$0.17}} & \rcell{tblGoodLt}{4.57 {\scriptsize $\pm$0.07}} & \rcell{tblGoodLt}{0.86 {\scriptsize $\pm$0.00}} & \rcell{tblNeutral}{0.86 {\scriptsize $\pm$0.00}} \\
Qwen 3 Next: NR {\tiny (3B-80B)} & MoE & 262k & \rcell{tblNeutral}{8.88 {\scriptsize $\pm$0.24}} & \rcell{tblGoodLt}{4.30 {\scriptsize $\pm$0.05}} & \rcell{tblGoodLt}{0.87 {\scriptsize $\pm$0.00}} & \rcell{tblGoodDk}{\textbf{0.88} {\scriptsize $\pm$0.00}} \\
Gemma 3 IT {\tiny (27B)} & Dense & 131k & \rcell{tblBadLt}{12.12 {\scriptsize $\pm$0.25}} & \rcell{tblGoodLt}{4.76 {\scriptsize $\pm$0.15}} & \rcell{tblNeutral}{0.77 {\scriptsize $\pm$0.01}} & \rcell{tblBadLt}{0.85 {\scriptsize $\pm$0.00}} \\
Llama 3.1 {\tiny (8B)} & Dense & 131k & \rcell{tblBadLt}{13.71 {\scriptsize $\pm$1.45}} & \rcell{tblNeutral}{8.30 {\scriptsize $\pm$1.22}} & \rcell{tblNeutral}{0.74 {\scriptsize $\pm$0.04}} & \rcell{tblBadLt}{0.81 {\scriptsize $\pm$0.01}} \\
\addlinespace[1.5pt]
\end{tabularx}
\end{tcolorbox}
\end{table*}


\begin{table*}[!tb]
\centering
\scriptsize
\caption{Item-wise MAE\,$\pm$\,std for I1--I10 with mean. Formatted as Table~\ref{tab:table2_total_scores}.}
\label{tab:table3_item_scores}
\begin{tcolorbox}[
  enhanced,
  boxrule=0.5pt,
  colframe=tblBorder,
  colback=white,
  arc=8pt,
  outer arc=8pt,
  left=3pt, right=3pt, top=3pt, bottom=3pt,
  boxsep=0pt,
  before upper={\arrayrulecolor{tblBorder}\renewcommand{\arraystretch}{1.25}\setlength{\tabcolsep}{2pt}},
]
\begin{tabularx}{\linewidth}{@{} >{\hsize=0.30\hsize\raggedright\arraybackslash}X >{\hsize=0.07\hsize\centering\arraybackslash}X >{\hsize=0.07\hsize\centering\arraybackslash}X >{\hsize=0.07\hsize\centering\arraybackslash}X >{\hsize=0.07\hsize\centering\arraybackslash}X >{\hsize=0.07\hsize\centering\arraybackslash}X >{\hsize=0.07\hsize\centering\arraybackslash}X >{\hsize=0.07\hsize\centering\arraybackslash}X >{\hsize=0.07\hsize\centering\arraybackslash}X >{\hsize=0.07\hsize\centering\arraybackslash}X >{\hsize=0.07\hsize\centering\arraybackslash}X !{\color{tblBorder}\vrule width 0.5pt} >{\hsize=0.07\hsize\centering\arraybackslash}X @{}}
\arrayrulecolor{tblBorder}
\rcell{tblNeutral}{\textsf{Model}\,{\scriptsize\textcolor{hdrSub}{(Size)}}} & \rcell{tblNeutral}{\textsf{I1}} & \rcell{tblNeutral}{\textsf{I2}} & \rcell{tblNeutral}{\textsf{I3}} & \rcell{tblNeutral}{\textsf{I4}} & \rcell{tblNeutral}{\textsf{I5}} & \rcell{tblNeutral}{\textsf{I6}} & \rcell{tblNeutral}{\textsf{I7}} & \rcell{tblNeutral}{\textsf{I8}} & \rcell{tblNeutral}{\textsf{I9}} & \rcell{tblNeutral}{\textsf{I10}} & \rcell{tblNeutral}{\textsf{Mean}} \\
\addlinespace[2pt]
\rowcolor{tblNeutral} \multicolumn{12}{@{}l@{}}{\small\textsc{Reasoning Models}} \\
\addlinespace[2pt]
Qwen 3 {\tiny (22B-235B)} & \rcell{tblNeutral}{\makecell{0.95 \\ {\tiny $\pm$ 0.03}}} & \rcell{tblNeutral}{\makecell{0.88 \\ {\tiny $\pm$ 0.03}}} & \rcell{tblNeutral}{\makecell{0.75 \\ {\tiny $\pm$ 0.03}}} & \rcell{tblNeutral}{\makecell{0.93 \\ {\tiny $\pm$ 0.02}}} & \rcell{tblGoodLt}{\makecell{0.40 \\ {\tiny $\pm$ 0.02}}} & \rcell{tblNeutral}{\makecell{0.88 \\ {\tiny $\pm$ 0.02}}} & \rcell{tblNeutral}{\makecell{0.75 \\ {\tiny $\pm$ 0.03}}} & \rcell{tblNeutral}{\makecell{0.78 \\ {\tiny $\pm$ 0.01}}} & \rcell{tblNeutral}{\makecell{0.66 \\ {\tiny $\pm$ 0.01}}} & \rcell{tblNeutral}{\makecell{0.64 \\ {\tiny $\pm$ 0.03}}} & \rcell{tblNeutral}{\makecell{0.76 \\ {\tiny $\pm$ 0.01}}} \\
GPT OSS 120B {\tiny (5B-117B)} & \rcell{tblNeutral}{\makecell{0.98 \\ {\tiny $\pm$ 0.01}}} & \rcell{tblNeutral}{\makecell{0.94 \\ {\tiny $\pm$ 0.01}}} & \rcell{tblNeutral}{\makecell{0.76 \\ {\tiny $\pm$ 0.02}}} & \rcell{tblNeutral}{\makecell{0.87 \\ {\tiny $\pm$ 0.02}}} & \rcell{tblGoodDk}{\makecell{\textbf{0.39} \\ {\tiny $\pm$ 0.01}}} & \rcell{tblNeutral}{\makecell{0.87 \\ {\tiny $\pm$ 0.02}}} & \rcell{tblNeutral}{\makecell{0.79 \\ {\tiny $\pm$ 0.02}}} & \rcell{tblNeutral}{\makecell{0.76 \\ {\tiny $\pm$ 0.03}}} & \rcell{tblNeutral}{\makecell{0.65 \\ {\tiny $\pm$ 0.02}}} & \rcell{tblNeutral}{\makecell{0.70 \\ {\tiny $\pm$ 0.04}}} & \rcell{tblNeutral}{\makecell{0.77 \\ {\tiny $\pm$ 0.02}}} \\
Magistral Small 2507 {\tiny (24B)} & \rcell{tblNeutral}{\makecell{1.04 \\ {\tiny $\pm$ 0.05}}} & \rcell{tblNeutral}{\makecell{0.94 \\ {\tiny $\pm$ 0.02}}} & \rcell{tblNeutral}{\makecell{0.80 \\ {\tiny $\pm$ 0.01}}} & \rcell{tblNeutral}{\makecell{\textbf{0.81} \\ {\tiny $\pm$ 0.00}}} & \rcell{tblGoodLt}{\makecell{0.44 \\ {\tiny $\pm$ 0.02}}} & \rcell{tblNeutral}{\makecell{0.85 \\ {\tiny $\pm$ 0.01}}} & \rcell{tblNeutral}{\makecell{0.75 \\ {\tiny $\pm$ 0.05}}} & \rcell{tblNeutral}{\makecell{0.78 \\ {\tiny $\pm$ 0.01}}} & \rcell{tblNeutral}{\makecell{0.73 \\ {\tiny $\pm$ 0.01}}} & \rcell{tblNeutral}{\makecell{0.60 \\ {\tiny $\pm$ 0.01}}} & \rcell{tblNeutral}{\makecell{0.77 \\ {\tiny $\pm$ 0.01}}} \\
DeepSeek R1 Qwen 2.5 {\tiny (32B)} & \rcell{tblNeutral}{\makecell{1.05 \\ {\tiny $\pm$ 0.01}}} & \rcell{tblNeutral}{\makecell{0.95 \\ {\tiny $\pm$ 0.02}}} & \rcell{tblNeutral}{\makecell{0.82 \\ {\tiny $\pm$ 0.03}}} & \rcell{tblNeutral}{\makecell{0.91 \\ {\tiny $\pm$ 0.05}}} & \rcell{tblGoodLt}{\makecell{0.47 \\ {\tiny $\pm$ 0.01}}} & \rcell{tblNeutral}{\makecell{0.96 \\ {\tiny $\pm$ 0.02}}} & \rcell{tblNeutral}{\makecell{0.74 \\ {\tiny $\pm$ 0.03}}} & \rcell{tblNeutral}{\makecell{0.80 \\ {\tiny $\pm$ 0.01}}} & \rcell{tblNeutral}{\makecell{0.73 \\ {\tiny $\pm$ 0.02}}} & \rcell{tblNeutral}{\makecell{0.67 \\ {\tiny $\pm$ 0.03}}} & \rcell{tblNeutral}{\makecell{0.81 \\ {\tiny $\pm$ 0.01}}} \\
DeepSeek R1 Llama 3.3 {\tiny (70B)} & \rcell{tblNeutral}{\makecell{1.06 \\ {\tiny $\pm$ 0.01}}} & \rcell{tblNeutral}{\makecell{0.92 \\ {\tiny $\pm$ 0.02}}} & \rcell{tblNeutral}{\makecell{0.78 \\ {\tiny $\pm$ 0.01}}} & \rcell{tblNeutral}{\makecell{0.92 \\ {\tiny $\pm$ 0.04}}} & \rcell{tblGoodLt}{\makecell{0.49 \\ {\tiny $\pm$ 0.02}}} & \rcell{tblNeutral}{\makecell{0.89 \\ {\tiny $\pm$ 0.01}}} & \rcell{tblNeutral}{\makecell{0.84 \\ {\tiny $\pm$ 0.01}}} & \rcell{tblNeutral}{\makecell{0.92 \\ {\tiny $\pm$ 0.02}}} & \rcell{tblNeutral}{\makecell{0.74 \\ {\tiny $\pm$ 0.01}}} & \rcell{tblNeutral}{\makecell{0.67 \\ {\tiny $\pm$ 0.02}}} & \rcell{tblNeutral}{\makecell{0.82 \\ {\tiny $\pm$ 0.00}}} \\
GPT OSS 20B {\tiny (3B-21B)} & \rcell{tblNeutral}{\makecell{1.07 \\ {\tiny $\pm$ 0.01}}} & \rcell{tblNeutral}{\makecell{0.98 \\ {\tiny $\pm$ 0.03}}} & \rcell{tblNeutral}{\makecell{0.85 \\ {\tiny $\pm$ 0.02}}} & \rcell{tblNeutral}{\makecell{0.96 \\ {\tiny $\pm$ 0.02}}} & \rcell{tblGoodLt}{\makecell{0.48 \\ {\tiny $\pm$ 0.02}}} & \rcell{tblNeutral}{\makecell{0.99 \\ {\tiny $\pm$ 0.02}}} & \rcell{tblNeutral}{\makecell{0.78 \\ {\tiny $\pm$ 0.02}}} & \rcell{tblNeutral}{\makecell{0.82 \\ {\tiny $\pm$ 0.02}}} & \rcell{tblNeutral}{\makecell{0.68 \\ {\tiny $\pm$ 0.00}}} & \rcell{tblNeutral}{\makecell{0.69 \\ {\tiny $\pm$ 0.03}}} & \rcell{tblNeutral}{\makecell{0.83 \\ {\tiny $\pm$ 0.01}}} \\
Qwen 3 {\tiny (32B)} & \rcell{tblNeutral}{\makecell{1.06 \\ {\tiny $\pm$ 0.02}}} & \rcell{tblNeutral}{\makecell{0.87 \\ {\tiny $\pm$ 0.02}}} & \rcell{tblNeutral}{\makecell{0.77 \\ {\tiny $\pm$ 0.02}}} & \rcell{tblNeutral}{\makecell{1.00 \\ {\tiny $\pm$ 0.02}}} & \rcell{tblGoodLt}{\makecell{0.53 \\ {\tiny $\pm$ 0.02}}} & \rcell{tblNeutral}{\makecell{0.93 \\ {\tiny $\pm$ 0.02}}} & \rcell{tblNeutral}{\makecell{0.81 \\ {\tiny $\pm$ 0.02}}} & \rcell{tblNeutral}{\makecell{0.89 \\ {\tiny $\pm$ 0.02}}} & \rcell{tblNeutral}{\makecell{0.69 \\ {\tiny $\pm$ 0.02}}} & \rcell{tblNeutral}{\makecell{0.82 \\ {\tiny $\pm$ 0.01}}} & \rcell{tblNeutral}{\makecell{0.84 \\ {\tiny $\pm$ 0.00}}} \\
QwQ {\tiny (32B)} & \rcell{tblNeutral}{\makecell{1.07 \\ {\tiny $\pm$ 0.01}}} & \rcell{tblNeutral}{\makecell{1.00 \\ {\tiny $\pm$ 0.02}}} & \rcell{tblNeutral}{\makecell{0.82 \\ {\tiny $\pm$ 0.03}}} & \rcell{tblNeutral}{\makecell{0.95 \\ {\tiny $\pm$ 0.02}}} & \rcell{tblGoodLt}{\makecell{0.56 \\ {\tiny $\pm$ 0.01}}} & \rcell{tblNeutral}{\makecell{1.03 \\ {\tiny $\pm$ 0.04}}} & \rcell{tblNeutral}{\makecell{0.79 \\ {\tiny $\pm$ 0.02}}} & \rcell{tblNeutral}{\makecell{0.88 \\ {\tiny $\pm$ 0.04}}} & \rcell{tblNeutral}{\makecell{0.72 \\ {\tiny $\pm$ 0.01}}} & \rcell{tblNeutral}{\makecell{0.78 \\ {\tiny $\pm$ 0.02}}} & \rcell{tblNeutral}{\makecell{0.86 \\ {\tiny $\pm$ 0.01}}} \\
Qwen 3 Next {\tiny (3B-80B)} & \rcell{tblNeutral}{\makecell{1.04 \\ {\tiny $\pm$ 0.02}}} & \rcell{tblNeutral}{\makecell{1.04 \\ {\tiny $\pm$ 0.02}}} & \rcell{tblNeutral}{\makecell{0.88 \\ {\tiny $\pm$ 0.01}}} & \rcell{tblNeutral}{\makecell{1.00 \\ {\tiny $\pm$ 0.01}}} & \rcell{tblGoodLt}{\makecell{0.53 \\ {\tiny $\pm$ 0.01}}} & \rcell{tblNeutral}{\makecell{0.99 \\ {\tiny $\pm$ 0.02}}} & \rcell{tblNeutral}{\makecell{1.02 \\ {\tiny $\pm$ 0.03}}} & \rcell{tblNeutral}{\makecell{0.98 \\ {\tiny $\pm$ 0.02}}} & \rcell{tblNeutral}{\makecell{0.76 \\ {\tiny $\pm$ 0.03}}} & \rcell{tblNeutral}{\makecell{0.69 \\ {\tiny $\pm$ 0.01}}} & \rcell{tblNeutral}{\makecell{0.89 \\ {\tiny $\pm$ 0.01}}} \\
Qwen 3 {\tiny (14B)} & \rcell{tblNeutral}{\makecell{1.16 \\ {\tiny $\pm$ 0.03}}} & \rcell{tblNeutral}{\makecell{1.00 \\ {\tiny $\pm$ 0.01}}} & \rcell{tblNeutral}{\makecell{0.91 \\ {\tiny $\pm$ 0.02}}} & \rcell{tblNeutral}{\makecell{1.10 \\ {\tiny $\pm$ 0.04}}} & \rcell{tblGoodLt}{\makecell{0.51 \\ {\tiny $\pm$ 0.01}}} & \rcell{tblNeutral}{\makecell{1.06 \\ {\tiny $\pm$ 0.03}}} & \rcell{tblNeutral}{\makecell{0.86 \\ {\tiny $\pm$ 0.01}}} & \rcell{tblNeutral}{\makecell{0.94 \\ {\tiny $\pm$ 0.03}}} & \rcell{tblNeutral}{\makecell{0.71 \\ {\tiny $\pm$ 0.02}}} & \rcell{tblNeutral}{\makecell{0.75 \\ {\tiny $\pm$ 0.03}}} & \rcell{tblNeutral}{\makecell{0.90 \\ {\tiny $\pm$ 0.01}}} \\
Qwen 3 {\tiny (8B)} & \rcell{tblNeutral}{\makecell{1.18 \\ {\tiny $\pm$ 0.04}}} & \rcell{tblNeutral}{\makecell{1.06 \\ {\tiny $\pm$ 0.00}}} & \rcell{tblNeutral}{\makecell{0.81 \\ {\tiny $\pm$ 0.01}}} & \rcell{tblBadLt}{\makecell{1.23 \\ {\tiny $\pm$ 0.03}}} & \rcell{tblNeutral}{\makecell{0.70 \\ {\tiny $\pm$ 0.04}}} & \rcell{tblNeutral}{\makecell{1.01 \\ {\tiny $\pm$ 0.02}}} & \rcell{tblNeutral}{\makecell{0.88 \\ {\tiny $\pm$ 0.03}}} & \rcell{tblBadLt}{\makecell{1.34 \\ {\tiny $\pm$ 0.03}}} & \rcell{tblNeutral}{\makecell{0.71 \\ {\tiny $\pm$ 0.02}}} & \rcell{tblNeutral}{\makecell{0.79 \\ {\tiny $\pm$ 0.03}}} & \rcell{tblNeutral}{\makecell{0.97 \\ {\tiny $\pm$ 0.01}}} \\
Qwen 3 {\tiny (3B-30B)} & \rcell{tblBadLt}{\makecell{1.25 \\ {\tiny $\pm$ 0.06}}} & \rcell{tblNeutral}{\makecell{1.08 \\ {\tiny $\pm$ 0.02}}} & \rcell{tblNeutral}{\makecell{1.09 \\ {\tiny $\pm$ 0.04}}} & \rcell{tblNeutral}{\makecell{1.17 \\ {\tiny $\pm$ 0.01}}} & \rcell{tblNeutral}{\makecell{0.62 \\ {\tiny $\pm$ 0.01}}} & \rcell{tblNeutral}{\makecell{1.19 \\ {\tiny $\pm$ 0.04}}} & \rcell{tblNeutral}{\makecell{1.09 \\ {\tiny $\pm$ 0.02}}} & \rcell{tblNeutral}{\makecell{1.07 \\ {\tiny $\pm$ 0.04}}} & \rcell{tblNeutral}{\makecell{0.75 \\ {\tiny $\pm$ 0.02}}} & \rcell{tblNeutral}{\makecell{0.85 \\ {\tiny $\pm$ 0.04}}} & \rcell{tblNeutral}{\makecell{1.02 \\ {\tiny $\pm$ 0.01}}} \\
Qwen 3 {\tiny (4B)} & \rcell{tblBadLt}{\makecell{1.27 \\ {\tiny $\pm$ 0.02}}} & \rcell{tblNeutral}{\makecell{1.09 \\ {\tiny $\pm$ 0.03}}} & \rcell{tblNeutral}{\makecell{0.95 \\ {\tiny $\pm$ 0.02}}} & \rcell{tblBadLt}{\makecell{1.20 \\ {\tiny $\pm$ 0.02}}} & \rcell{tblNeutral}{\makecell{0.80 \\ {\tiny $\pm$ 0.02}}} & \rcell{tblBadLt}{\makecell{1.27 \\ {\tiny $\pm$ 0.06}}} & \rcell{tblNeutral}{\makecell{1.04 \\ {\tiny $\pm$ 0.01}}} & \rcell{tblBadLt}{\makecell{1.69 \\ {\tiny $\pm$ 0.03}}} & \rcell{tblNeutral}{\makecell{0.81 \\ {\tiny $\pm$ 0.01}}} & \rcell{tblNeutral}{\makecell{0.94 \\ {\tiny $\pm$ 0.01}}} & \rcell{tblNeutral}{\makecell{1.11 \\ {\tiny $\pm$ 0.00}}} \\
Qwen 3 {\tiny (1.7B)} & \rcell{tblBadLt}{\makecell{1.68 \\ {\tiny $\pm$ 0.01}}} & \rcell{tblBadDk}{\makecell{1.51 \\ {\tiny $\pm$ 0.02}}} & \rcell{tblNeutral}{\makecell{1.16 \\ {\tiny $\pm$ 0.01}}} & \rcell{tblBadLt}{\makecell{1.43 \\ {\tiny $\pm$ 0.02}}} & \rcell{tblNeutral}{\makecell{0.97 \\ {\tiny $\pm$ 0.05}}} & \rcell{tblBadLt}{\makecell{1.28 \\ {\tiny $\pm$ 0.04}}} & \rcell{tblBadLt}{\makecell{1.23 \\ {\tiny $\pm$ 0.06}}} & \rcell{tblBadDk}{\makecell{2.45 \\ {\tiny $\pm$ 0.12}}} & \rcell{tblNeutral}{\makecell{0.93 \\ {\tiny $\pm$ 0.02}}} & \rcell{tblNeutral}{\makecell{0.94 \\ {\tiny $\pm$ 0.04}}} & \rcell{tblBadLt}{\makecell{1.36 \\ {\tiny $\pm$ 0.01}}} \\
DeepSeek R1 Llama 3.1 {\tiny (8B)} & \rcell{tblBadLt}{\makecell{1.53 \\ {\tiny $\pm$ 0.01}}} & \rcell{tblBadLt}{\makecell{1.41 \\ {\tiny $\pm$ 0.05}}} & \rcell{tblNeutral}{\makecell{0.96 \\ {\tiny $\pm$ 0.06}}} & \rcell{tblBadLt}{\makecell{1.61 \\ {\tiny $\pm$ 0.10}}} & \rcell{tblBadLt}{\makecell{1.65 \\ {\tiny $\pm$ 0.13}}} & \rcell{tblBadLt}{\makecell{1.29 \\ {\tiny $\pm$ 0.06}}} & \rcell{tblBadDk}{\makecell{1.73 \\ {\tiny $\pm$ 0.08}}} & \rcell{tblBadLt}{\makecell{2.20 \\ {\tiny $\pm$ 0.14}}} & \rcell{tblNeutral}{\makecell{1.06 \\ {\tiny $\pm$ 0.06}}} & \rcell{tblBadLt}{\makecell{1.39 \\ {\tiny $\pm$ 0.17}}} & \rcell{tblBadLt}{\makecell{1.48 \\ {\tiny $\pm$ 0.07}}} \\
Qwen 3 {\tiny (0.6B)} & \rcell{tblBadDk}{\makecell{1.78 \\ {\tiny $\pm$ 0.04}}} & \rcell{tblBadLt}{\makecell{1.47 \\ {\tiny $\pm$ 0.02}}} & \rcell{tblBadLt}{\makecell{1.32 \\ {\tiny $\pm$ 0.03}}} & \rcell{tblBadDk}{\makecell{2.28 \\ {\tiny $\pm$ 0.05}}} & \rcell{tblBadDk}{\makecell{1.72 \\ {\tiny $\pm$ 0.05}}} & \rcell{tblBadDk}{\makecell{1.53 \\ {\tiny $\pm$ 0.01}}} & \rcell{tblBadLt}{\makecell{1.72 \\ {\tiny $\pm$ 0.05}}} & \rcell{tblBadLt}{\makecell{1.84 \\ {\tiny $\pm$ 0.07}}} & \rcell{tblBadDk}{\makecell{1.37 \\ {\tiny $\pm$ 0.03}}} & \rcell{tblBadDk}{\makecell{1.50 \\ {\tiny $\pm$ 0.03}}} & \rcell{tblBadDk}{\makecell{1.65 \\ {\tiny $\pm$ 0.01}}} \\
\addlinespace[2pt]
\rowcolor{tblNeutral} \multicolumn{12}{@{}l@{}}{\small\textsc{Non-Reasoning Models}} \\
\addlinespace[2pt]
Qwen 2.5 {\tiny (72B)} & \rcell{tblNeutral}{\makecell{0.94 \\ {\tiny $\pm$ 0.02}}} & \rcell{tblNeutral}{\makecell{\textbf{0.77} \\ {\tiny $\pm$ 0.01}}} & \rcell{tblNeutral}{\makecell{\textbf{0.66} \\ {\tiny $\pm$ 0.02}}} & \rcell{tblNeutral}{\makecell{0.92 \\ {\tiny $\pm$ 0.01}}} & \rcell{tblGoodLt}{\makecell{0.50 \\ {\tiny $\pm$ 0.02}}} & \rcell{tblNeutral}{\makecell{\textbf{0.80} \\ {\tiny $\pm$ 0.02}}} & \rcell{tblNeutral}{\makecell{\textbf{0.71} \\ {\tiny $\pm$ 0.01}}} & \rcell{tblNeutral}{\makecell{0.89 \\ {\tiny $\pm$ 0.01}}} & \rcell{tblNeutral}{\makecell{0.65 \\ {\tiny $\pm$ 0.01}}} & \rcell{tblGoodLt}{\makecell{0.58 \\ {\tiny $\pm$ 0.01}}} & \rcell{tblNeutral}{\makecell{\textbf{0.74} \\ {\tiny $\pm$ 0.00}}} \\
Llama 4 Scout {\tiny (17B-109B)} & \rcell{tblNeutral}{\makecell{1.06 \\ {\tiny $\pm$ 0.01}}} & \rcell{tblNeutral}{\makecell{0.89 \\ {\tiny $\pm$ 0.01}}} & \rcell{tblNeutral}{\makecell{0.76 \\ {\tiny $\pm$ 0.01}}} & \rcell{tblNeutral}{\makecell{0.83 \\ {\tiny $\pm$ 0.01}}} & \rcell{tblGoodLt}{\makecell{0.42 \\ {\tiny $\pm$ 0.01}}} & \rcell{tblNeutral}{\makecell{0.83 \\ {\tiny $\pm$ 0.01}}} & \rcell{tblNeutral}{\makecell{0.73 \\ {\tiny $\pm$ 0.00}}} & \rcell{tblNeutral}{\makecell{\textbf{0.76} \\ {\tiny $\pm$ 0.00}}} & \rcell{tblNeutral}{\makecell{0.66 \\ {\tiny $\pm$ 0.01}}} & \rcell{tblGoodLt}{\makecell{0.60 \\ {\tiny $\pm$ 0.01}}} & \rcell{tblNeutral}{\makecell{0.75 \\ {\tiny $\pm$ 0.00}}} \\
Qwen 2.5 {\tiny (14B): 1M} & \rcell{tblNeutral}{\makecell{\textbf{0.89} \\ {\tiny $\pm$ 0.01}}} & \rcell{tblNeutral}{\makecell{0.90 \\ {\tiny $\pm$ 0.03}}} & \rcell{tblNeutral}{\makecell{0.80 \\ {\tiny $\pm$ 0.03}}} & \rcell{tblNeutral}{\makecell{0.89 \\ {\tiny $\pm$ 0.02}}} & \rcell{tblGoodLt}{\makecell{0.48 \\ {\tiny $\pm$ 0.01}}} & \rcell{tblNeutral}{\makecell{0.85 \\ {\tiny $\pm$ 0.02}}} & \rcell{tblNeutral}{\makecell{0.78 \\ {\tiny $\pm$ 0.02}}} & \rcell{tblNeutral}{\makecell{0.97 \\ {\tiny $\pm$ 0.02}}} & \rcell{tblNeutral}{\makecell{0.71 \\ {\tiny $\pm$ 0.02}}} & \rcell{tblGoodDk}{\makecell{\textbf{0.57} \\ {\tiny $\pm$ 0.02}}} & \rcell{tblNeutral}{\makecell{0.78 \\ {\tiny $\pm$ 0.01}}} \\
Llama 4 Maverick {\tiny (17B-400B)} & \rcell{tblNeutral}{\makecell{0.96 \\ {\tiny $\pm$ 0.00}}} & \rcell{tblNeutral}{\makecell{0.83 \\ {\tiny $\pm$ 0.01}}} & \rcell{tblNeutral}{\makecell{0.75 \\ {\tiny $\pm$ 0.01}}} & \rcell{tblNeutral}{\makecell{0.88 \\ {\tiny $\pm$ 0.01}}} & \rcell{tblGoodLt}{\makecell{0.42 \\ {\tiny $\pm$ 0.01}}} & \rcell{tblNeutral}{\makecell{0.91 \\ {\tiny $\pm$ 0.02}}} & \rcell{tblNeutral}{\makecell{0.92 \\ {\tiny $\pm$ 0.01}}} & \rcell{tblNeutral}{\makecell{1.06 \\ {\tiny $\pm$ 0.02}}} & \rcell{tblNeutral}{\makecell{\textbf{0.63} \\ {\tiny $\pm$ 0.00}}} & \rcell{tblGoodLt}{\makecell{0.59 \\ {\tiny $\pm$ 0.01}}} & \rcell{tblNeutral}{\makecell{0.80 \\ {\tiny $\pm$ 0.00}}} \\
Llama 3.3 {\tiny (70B)} & \rcell{tblNeutral}{\makecell{1.07 \\ {\tiny $\pm$ 0.01}}} & \rcell{tblNeutral}{\makecell{0.83 \\ {\tiny $\pm$ 0.01}}} & \rcell{tblNeutral}{\makecell{0.83 \\ {\tiny $\pm$ 0.01}}} & \rcell{tblNeutral}{\makecell{0.92 \\ {\tiny $\pm$ 0.00}}} & \rcell{tblGoodLt}{\makecell{0.52 \\ {\tiny $\pm$ 0.02}}} & \rcell{tblNeutral}{\makecell{0.82 \\ {\tiny $\pm$ 0.00}}} & \rcell{tblNeutral}{\makecell{0.79 \\ {\tiny $\pm$ 0.01}}} & \rcell{tblNeutral}{\makecell{0.85 \\ {\tiny $\pm$ 0.01}}} & \rcell{tblNeutral}{\makecell{0.78 \\ {\tiny $\pm$ 0.01}}} & \rcell{tblNeutral}{\makecell{0.66 \\ {\tiny $\pm$ 0.01}}} & \rcell{tblNeutral}{\makecell{0.81 \\ {\tiny $\pm$ 0.01}}} \\
Gemma 3 IT {\tiny (27B)} & \rcell{tblNeutral}{\makecell{1.08 \\ {\tiny $\pm$ 0.00}}} & \rcell{tblNeutral}{\makecell{0.92 \\ {\tiny $\pm$ 0.01}}} & \rcell{tblNeutral}{\makecell{0.99 \\ {\tiny $\pm$ 0.02}}} & \rcell{tblNeutral}{\makecell{0.96 \\ {\tiny $\pm$ 0.01}}} & \rcell{tblNeutral}{\makecell{0.62 \\ {\tiny $\pm$ 0.01}}} & \rcell{tblNeutral}{\makecell{0.83 \\ {\tiny $\pm$ 0.01}}} & \rcell{tblNeutral}{\makecell{0.81 \\ {\tiny $\pm$ 0.03}}} & \rcell{tblNeutral}{\makecell{1.06 \\ {\tiny $\pm$ 0.03}}} & \rcell{tblNeutral}{\makecell{0.82 \\ {\tiny $\pm$ 0.02}}} & \rcell{tblNeutral}{\makecell{0.69 \\ {\tiny $\pm$ 0.01}}} & \rcell{tblNeutral}{\makecell{0.88 \\ {\tiny $\pm$ 0.01}}} \\
Qwen 3 Next: NR {\tiny (3B-80B)} & \rcell{tblNeutral}{\makecell{1.10 \\ {\tiny $\pm$ 0.01}}} & \rcell{tblNeutral}{\makecell{1.04 \\ {\tiny $\pm$ 0.01}}} & \rcell{tblNeutral}{\makecell{0.94 \\ {\tiny $\pm$ 0.00}}} & \rcell{tblNeutral}{\makecell{0.95 \\ {\tiny $\pm$ 0.01}}} & \rcell{tblNeutral}{\makecell{0.63 \\ {\tiny $\pm$ 0.00}}} & \rcell{tblNeutral}{\makecell{0.88 \\ {\tiny $\pm$ 0.00}}} & \rcell{tblNeutral}{\makecell{0.92 \\ {\tiny $\pm$ 0.02}}} & \rcell{tblNeutral}{\makecell{0.87 \\ {\tiny $\pm$ 0.01}}} & \rcell{tblNeutral}{\makecell{0.83 \\ {\tiny $\pm$ 0.01}}} & \rcell{tblNeutral}{\makecell{0.62 \\ {\tiny $\pm$ 0.01}}} & \rcell{tblNeutral}{\makecell{0.88 \\ {\tiny $\pm$ 0.00}}} \\
Qwen 2.5 {\tiny (7B): 1M} & \rcell{tblNeutral}{\makecell{1.08 \\ {\tiny $\pm$ 0.03}}} & \rcell{tblNeutral}{\makecell{0.99 \\ {\tiny $\pm$ 0.04}}} & \rcell{tblNeutral}{\makecell{0.78 \\ {\tiny $\pm$ 0.01}}} & \rcell{tblNeutral}{\makecell{0.92 \\ {\tiny $\pm$ 0.01}}} & \rcell{tblGoodLt}{\makecell{0.59 \\ {\tiny $\pm$ 0.02}}} & \rcell{tblNeutral}{\makecell{0.93 \\ {\tiny $\pm$ 0.02}}} & \rcell{tblNeutral}{\makecell{1.08 \\ {\tiny $\pm$ 0.04}}} & \rcell{tblNeutral}{\makecell{0.96 \\ {\tiny $\pm$ 0.02}}} & \rcell{tblNeutral}{\makecell{1.00 \\ {\tiny $\pm$ 0.03}}} & \rcell{tblNeutral}{\makecell{0.67 \\ {\tiny $\pm$ 0.04}}} & \rcell{tblNeutral}{\makecell{0.90 \\ {\tiny $\pm$ 0.01}}} \\
Llama 3.1 {\tiny (8B)} & \rcell{tblBadLt}{\makecell{1.38 \\ {\tiny $\pm$ 0.06}}} & \rcell{tblNeutral}{\makecell{1.15 \\ {\tiny $\pm$ 0.12}}} & \rcell{tblBadDk}{\makecell{1.34 \\ {\tiny $\pm$ 0.07}}} & \rcell{tblBadLt}{\makecell{1.61 \\ {\tiny $\pm$ 0.15}}} & \rcell{tblNeutral}{\makecell{1.19 \\ {\tiny $\pm$ 0.22}}} & \rcell{tblNeutral}{\makecell{1.07 \\ {\tiny $\pm$ 0.07}}} & \rcell{tblBadLt}{\makecell{1.57 \\ {\tiny $\pm$ 0.10}}} & \rcell{tblBadLt}{\makecell{1.46 \\ {\tiny $\pm$ 0.08}}} & \rcell{tblBadLt}{\makecell{1.23 \\ {\tiny $\pm$ 0.08}}} & \rcell{tblNeutral}{\makecell{1.05 \\ {\tiny $\pm$ 0.01}}} & \rcell{tblBadLt}{\makecell{1.30 \\ {\tiny $\pm$ 0.09}}} \\
\addlinespace[2pt]
\end{tabularx}
\end{tcolorbox}
\end{table*}


\begin{table*}[!tb]
\centering
\scriptsize
\caption{Invalid output counts (mean across runs) per MADRS item and total. Lower is better. Formatted as Table~\ref{tab:table2_total_scores}.}
\label{tab:invalid_outputs}
\begin{tcolorbox}[
  enhanced,
  boxrule=0.5pt,
  colframe=tblBorder,
  colback=white,
  arc=8pt,
  outer arc=8pt,
  left=3pt, right=3pt, top=3pt, bottom=3pt,
  boxsep=0pt,
  before upper={\arrayrulecolor{tblBorder}\renewcommand{\arraystretch}{1.25}\setlength{\tabcolsep}{2pt}},
]
\begin{tabularx}{\linewidth}{@{} >{\hsize=0.30\hsize\raggedright\arraybackslash}X >{\hsize=0.07\hsize\centering\arraybackslash}X >{\hsize=0.07\hsize\centering\arraybackslash}X >{\hsize=0.07\hsize\centering\arraybackslash}X >{\hsize=0.07\hsize\centering\arraybackslash}X >{\hsize=0.07\hsize\centering\arraybackslash}X >{\hsize=0.07\hsize\centering\arraybackslash}X >{\hsize=0.07\hsize\centering\arraybackslash}X >{\hsize=0.07\hsize\centering\arraybackslash}X >{\hsize=0.07\hsize\centering\arraybackslash}X >{\hsize=0.07\hsize\centering\arraybackslash}X !{\color{tblBorder}\vrule width 0.5pt} >{\hsize=0.07\hsize\centering\arraybackslash}X @{}}
\arrayrulecolor{tblBorder}
\rcell{tblNeutral}{\textsf{Model}\,{\scriptsize\textcolor{hdrSub}{(Size)}}} & \rcell{tblNeutral}{\textsf{I1}} & \rcell{tblNeutral}{\textsf{I2}} & \rcell{tblNeutral}{\textsf{I3}} & \rcell{tblNeutral}{\textsf{I4}} & \rcell{tblNeutral}{\textsf{I5}} & \rcell{tblNeutral}{\textsf{I6}} & \rcell{tblNeutral}{\textsf{I7}} & \rcell{tblNeutral}{\textsf{I8}} & \rcell{tblNeutral}{\textsf{I9}} & \rcell{tblNeutral}{\textsf{I10}} & \rcell{tblNeutral}{\textsf{Total}} \\
\addlinespace[2pt]
\rowcolor{tblNeutral} \multicolumn{12}{@{}l@{}}{\small\textsc{Reasoning Models}} \\
\addlinespace[2pt]
Magistral Small 2507 {\tiny (24B)} & 0 & 0 & 0 & 0 & 0 & 0 & 0 & 0 & 0 & 0 & 0 \\
GPT OSS 120B {\tiny (5B-117B)} & 0 & 0 & \rcell{tblBadLt}{0.3} & 0 & \rcell{tblBadLt}{0.3} & 0 & 0 & 0 & 0 & 0 & \rcell{tblBadLt}{0.7} \\
DeepSeek R1 Llama 3.3 {\tiny (70B)} & 0 & 0 & \rcell{tblBadLt}{0.7} & 0 & 0 & 0 & 0 & 0 & 0 & \rcell{tblBadLt}{0.3} & \rcell{tblBadLt}{1.0} \\
GPT OSS 20B {\tiny (3B-21B)} & 0 & 0 & 0 & \rcell{tblBadLt}{0.3} & 0 & 0 & 0 & \rcell{tblBadLt}{0.7} & 0 & 0 & \rcell{tblBadLt}{1.0} \\
DeepSeek R1 Llama 3.1 {\tiny (8B)} & \rcell{tblBadLt}{0.3} & 0 & 0 & \rcell{tblBadLt}{0.7} & \rcell{tblBadLt}{0.3} & 0 & 0 & 0 & 0 & \rcell{tblBadLt}{0.3} & \rcell{tblBadLt}{1.7} \\
DeepSeek R1 Qwen 2.5 {\tiny (32B)} & \rcell{tblBadLt}{0.3} & \rcell{tblBadLt}{0.3} & \rcell{tblBadLt}{0.3} & \rcell{tblBadLt}{1.0} & 0 & \rcell{tblBadLt}{0.7} & 0 & 0 & \rcell{tblBadLt}{0.3} & 0 & \rcell{tblBadLt}{3.0} \\
Qwen 3 Next {\tiny (3B-80B)} & \rcell{tblBadLt}{0.3} & 0 & \rcell{tblBadLt}{1.3} & 0 & \rcell{tblBadLt}{0.7} & \rcell{tblBadLt}{0.3} & 0 & \rcell{tblBadLt}{0.7} & 0 & 0 & \rcell{tblBadLt}{3.3} \\
Qwen 3 {\tiny (32B)} & \rcell{tblBadLt}{1.7} & 0 & 0 & 0 & \rcell{tblBadLt}{0.7} & \rcell{tblBadLt}{0.7} & 0 & \rcell{tblBadLt}{1.3} & 0 & \rcell{tblBadLt}{0.3} & \rcell{tblBadLt}{4.7} \\
Qwen 3 {\tiny (8B)} & \rcell{tblBadLt}{0.7} & \rcell{tblBadLt}{1.0} & \rcell{tblBadLt}{1.0} & \rcell{tblBadLt}{1.0} & \rcell{tblBadLt}{3.0} & \rcell{tblBadLt}{0.3} & \rcell{tblBadLt}{1.0} & \rcell{tblBadLt}{0.7} & \rcell{tblBadLt}{0.3} & \rcell{tblBadLt}{1.0} & \rcell{tblBadDk}{10.0} \\
Qwen 3 {\tiny (22B-235B)} & \rcell{tblBadLt}{2.7} & \rcell{tblBadLt}{2.3} & \rcell{tblBadLt}{0.7} & \rcell{tblBadLt}{1.0} & \rcell{tblBadLt}{1.3} & 0 & \rcell{tblBadLt}{4.0} & \rcell{tblBadLt}{1.0} & \rcell{tblBadLt}{0.3} & \rcell{tblBadLt}{3.7} & \rcell{tblBadDk}{17.0} \\
Qwen 3 {\tiny (14B)} & \rcell{tblBadLt}{2.3} & \rcell{tblBadLt}{2.3} & \rcell{tblBadLt}{1.7} & \rcell{tblBadLt}{2.0} & \rcell{tblBadLt}{2.0} & \rcell{tblBadLt}{1.7} & \rcell{tblBadLt}{2.3} & \rcell{tblBadLt}{1.7} & 0 & \rcell{tblBadLt}{1.3} & \rcell{tblBadDk}{17.3} \\
Qwen 3 {\tiny (4B)} & \rcell{tblBadLt}{1.7} & \rcell{tblBadLt}{1.3} & \rcell{tblBadLt}{3.7} & \rcell{tblBadLt}{3.3} & \rcell{tblBadLt}{4.3} & \rcell{tblBadDk}{5.0} & \rcell{tblBadLt}{2.3} & \rcell{tblBadLt}{4.0} & \rcell{tblBadLt}{0.7} & \rcell{tblBadLt}{3.7} & \rcell{tblBadDk}{30.0} \\
QwQ {\tiny (32B)} & \rcell{tblBadLt}{2.7} & \rcell{tblBadLt}{3.7} & \rcell{tblBadLt}{4.0} & \rcell{tblBadLt}{4.7} & \rcell{tblBadLt}{4.7} & \rcell{tblBadLt}{3.0} & \rcell{tblBadLt}{4.0} & \rcell{tblBadLt}{3.0} & 0 & \rcell{tblBadDk}{5.0} & \rcell{tblBadDk}{34.7} \\
Qwen 3 {\tiny (0.6B)} & \rcell{tblBadDk}{7.7} & \rcell{tblBadDk}{7.0} & \rcell{tblBadDk}{9.7} & \rcell{tblBadDk}{9.0} & \rcell{tblBadDk}{5.7} & \rcell{tblBadDk}{9.3} & \rcell{tblBadDk}{9.3} & \rcell{tblBadDk}{9.3} & \rcell{tblBadLt}{0.7} & \rcell{tblBadDk}{8.7} & \rcell{tblBadDk}{76.3} \\
Qwen 3 {\tiny (1.7B)} & \rcell{tblBadDk}{8.3} & \rcell{tblBadDk}{5.3} & \rcell{tblBadDk}{10.7} & \rcell{tblBadDk}{8.3} & \rcell{tblBadDk}{12.0} & \rcell{tblBadDk}{11.0} & \rcell{tblBadDk}{8.0} & \rcell{tblBadDk}{8.7} & \rcell{tblBadLt}{1.3} & \rcell{tblBadDk}{6.3} & \rcell{tblBadDk}{80.0} \\
Qwen 3 {\tiny (3B-30B)} & \rcell{tblBadDk}{20.3} & \rcell{tblBadDk}{22.3} & \rcell{tblBadDk}{22.3} & \rcell{tblBadDk}{9.0} & \rcell{tblBadDk}{18.7} & \rcell{tblBadDk}{17.3} & \rcell{tblBadDk}{24.3} & \rcell{tblBadDk}{21.7} & 0 & \rcell{tblBadDk}{19.7} & \rcell{tblBadDk}{175.7} \\
\addlinespace[2pt]
\rowcolor{tblNeutral} \multicolumn{12}{@{}l@{}}{\small\textsc{Non-Reasoning Models}} \\
\addlinespace[2pt]
\rowcolor{tblGoodLt}\multicolumn{12}{@{}c@{}}{\textit{All 9 non-reasoning models produced 0 invalid outputs.}} \\
\addlinespace[2pt]
\end{tabularx}
\end{tcolorbox}
\end{table*}

\subsection{Item-then-Sum (\textsc{ItS}) vs.\ Direct Total Score (\textsc{DTS})}
\label{ssec:total-performance}

\Cref{tab:table2_total_scores} reveals a striking pattern: \textsc{ItS} aggregation substantially reduces total MAE for most models. Standard Qwen~2.5 (72B) improves from $5.53$ (\textsc{DTS}) to $3.80$ (\textsc{ItS}, $31$\% reduction); reasoning-augmented models show even larger improvements, with GPT~OSS~120B transitioning from $6.80$ to $3.78$ ($44$\% reduction) and QwQ (32B) dramatically improving from $21.09$ to $4.03$ ($81$\% reduction). Most models achieve total MAE in the acceptable band ($<6$) after aggregation; however, some low-parameter models (e.g., DeepSeek R1 Llama 3.1 8B with sum MAE of $11.84$, and Llama 3.1 8B with $8.30$) remain in the moderate or substantial error range, indicating fundamental capacity limitations below certain parameter thresholds.

Critically, reasoning-augmented models' internal aggregation—even when they explicitly sum items in their reasoning traces—remains less accurate than post-hoc item summation. Even when models show step-by-step arithmetic, their \textsc{DTS} predictions remain systematically inferior to mechanically summed item-level outputs, indicating fundamental limits of \textsc{DTS} prediction regardless of reasoning capability.

Output validity further compounds these capacity limitations: \Cref{tab:invalid_outputs} shows that reasoning-augmented models produce substantially more invalid total-score outputs than item-level ones (e.g., QwQ 32B averages $34.7$ invalid total outputs vs.\ $2.7$--$5.0$ per item), as the longer transcripts used in \textsc{DTS} prediction leave less generation budget for the reasoning trace, increasing the likelihood that the model fails to produce a score. This asymmetry partly explains the outsized \textsc{DTS}--\textsc{ItS} gap for reasoning models.

Screening performance (F$_1$ for total $\geq 20$) shows considerable variation: top-performing models achieve F$_1$ scores of $0.86$--$0.88$, while smaller models range from $0.57$ to $0.77$. For most mid-to-large scale models, differences between \textsc{DTS} and \textsc{ItS} approaches are modest ($\Delta < 0.05$), though some small models show larger improvements with aggregation (e.g., Qwen 3 1.7B: $0.57 \rightarrow 0.78$, $\Delta = +0.21$). The distinction between reasoning-augmented and standard models is minimal for screening tasks, with both architectures achieving comparable F$_1$ scores at similar parameter scales.

\begin{figure*}[t]
  \centering
  \includegraphics[width=\linewidth]{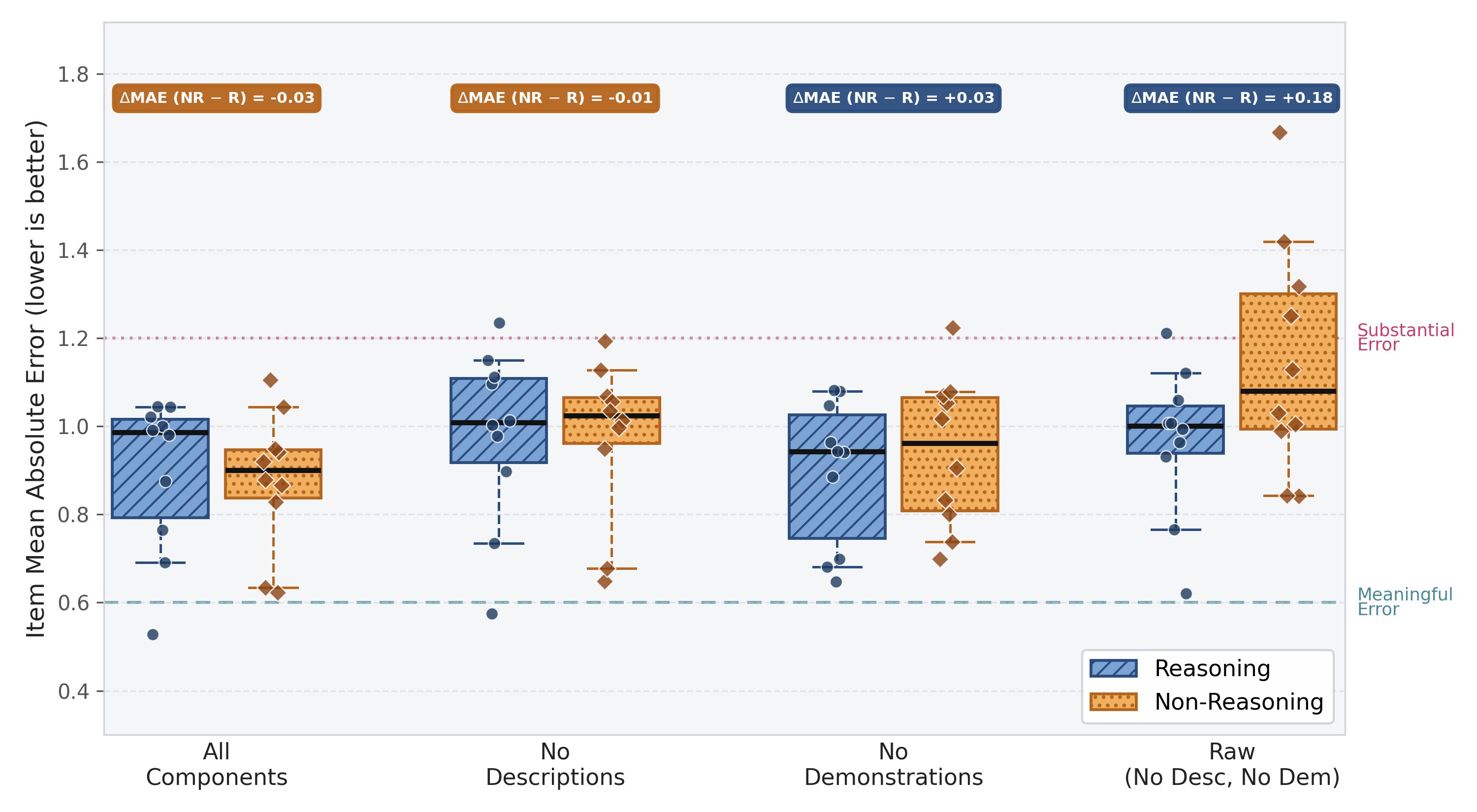}
  \caption{\textbf{Prompt ablation: descriptive \& demonstrative cues, reasoning vs. no-reasoning.}
  Mean Item MAE for Reasoning-Augmented (R) vs. Standard (NR) models across four prompt configurations. Values above each bar pair show $\Delta=\text{Standard}-\text{Reasoning}$; positive values indicate a reasoning advantage.}
  \label{fig:ablation-readable}
\end{figure*}
\subsection{Item-wise Performance}
\label{ssec:item-performance}

\Cref{tab:table3_item_scores} presents item-level MAE across all models, with performance categorized into acceptable ($<0.6$), moderate ($0.6$--$1.2$), and substantial ($\geq 1.2$) error bands. Among standard models, Qwen~2.5 (72B) achieves the lowest average MAE ($0.74$), with \emph{Reduced Appetite} ($0.50$) and \emph{Suicidal Thoughts} ($0.58$) falling in the acceptable band and remaining items in the moderate range ($0.65$--$0.94$). Llama~4~Scout attains comparable performance (mean $0.75$), again placing \emph{Reduced Appetite} at $0.42$ (acceptable error). Reasoning-augmented models show similar patterns: GPT~OSS~120B achieves mean MAE of $0.77$, and Qwen~3 (235B) achieves $0.76$, with \emph{Reduced Appetite} consistently in the acceptable band ($0.39$ and $0.40$, respectively). Notably, mid-scale models approach large-model performance: Qwen~2.5 (14B; 1M context) achieves mean MAE of $0.78$ with acceptable error on physiological symptom items.

In contrast, models with $\leq 8$B parameters exhibit multiple cells in the substantial band ($\geq 1.2$), with Llama~3.1 (8B) and Qwen~3 (0.6B--1.7B) frequently exceeding MAE of $1.3$ across several items. Comparing architectures, standard models demonstrate superior average item-level performance (mean MAE $0.87$) compared to reasoning-augmented models (mean MAE $0.99$), suggesting that explicit reasoning traces do not uniformly improve accuracy at this task granularity. 

Part of the accuracy gap between architectures is attributable to output validity: as \Cref{tab:invalid_outputs} shows, all nine non-reasoning models produced zero invalid outputs, whereas reasoning-augmented models—particularly at smaller scales—frequently exhaust their generation budget on reasoning traces before emitting a parseable prediction (e.g., Qwen~3 0.6B: $76.3$; 1.7B: $80.0$; 3B--30B MoE: $175.7$ invalid outputs on the total-score task).

Item-specific patterns reveal consistent difficulty hierarchies: \emph{Reduced Appetite} (Item 5) achieves acceptable error ($<0.6$) across all models regardless of architecture or scale, likely reflecting its concrete behavioral nature. Conversely, smaller reasoning-augmented models struggle particularly with \emph{Lassitude} (Item 7) and \emph{Inability to Feel} (Item 8), with MAE reaching $2.45$ for the smallest variant (Qwen~3 1.7B) on Item 8, indicating that affective-behavioral items pose unique challenges for low-capacity reasoning systems. The persistent difficulty of \emph{Apparent Sadness} (Item 1) across both architectures—despite interview questions specifically probing what others observe about the patient's appearance—suggests an inherent limitation of text-only assessment: this item fundamentally requires visual and paralinguistic observations that necessitate multimodal integration.

\begin{figure*}[t]
  \centering
    \centering
    \includegraphics[width=\linewidth]{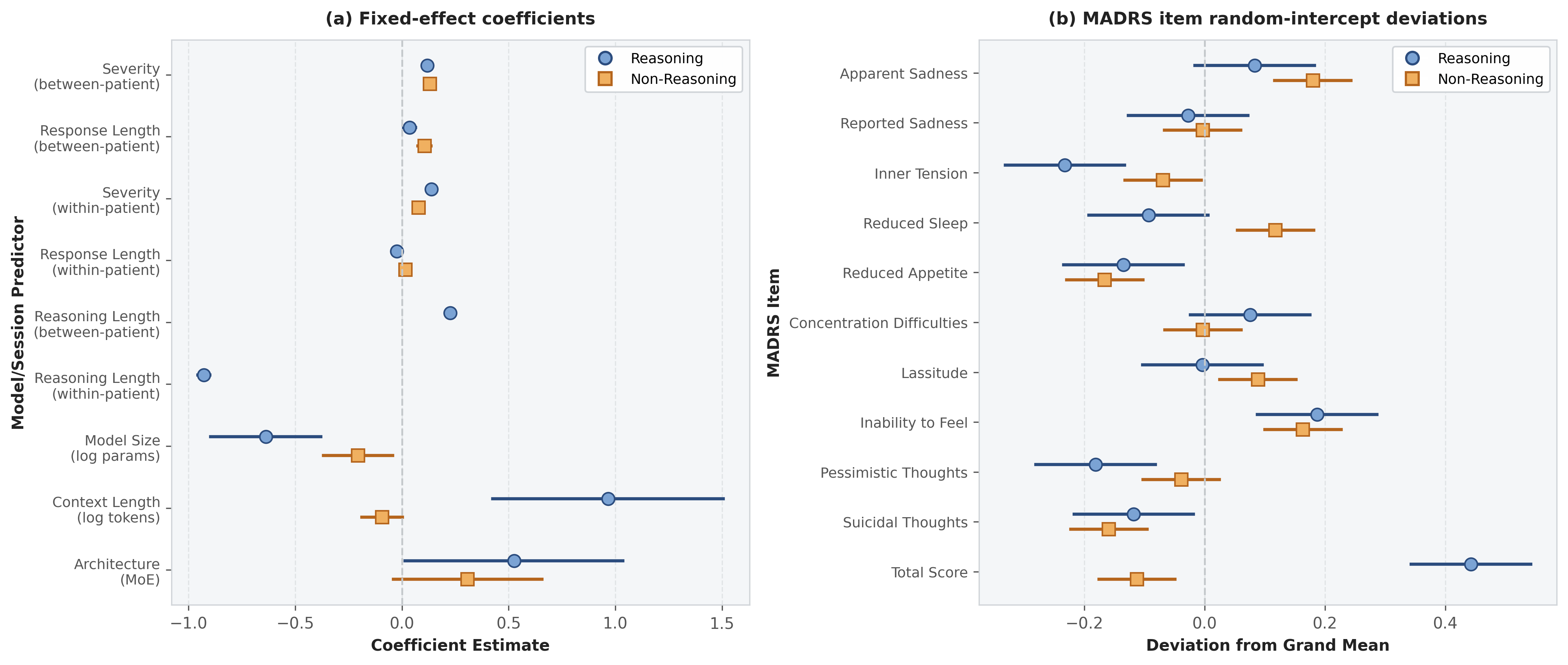}
\caption{\textbf{Mixed-effects analysis (task-level, seed-averaged).} Blue = reasoning; red = standard. 
(a) Fixed-effect estimates (points: coefficients; bars: 95\% CI). Longer \emph{within-patient} reasoning is linked to lower error, while patients/models with \emph{typically} longer reasoning show higher error. Larger models reduce error; MoE tends to increase it. 
(b) Item random-intercept deviations (BLUPs) highlight item-specific difficulty (e.g., \textit{Inability to Feel} $>$0; \textit{Inner Tension} $<$0).}
  \label{fig:items-main}
\end{figure*}

\subsection{Prompt Ablation: When Does Reasoning Help?}
\label{ssec:ablation}

Using \textbf{Qwen~3~Next (80B; MoE)} with and without reasoning augmentation, we ablated descriptive cues and demonstrative cues across four configurations. \Cref{fig:ablation-readable} presents paired comparisons (reasoning-augmented in blue, standard in red), with $\Delta=\text{Standard}-\text{Reasoning}$ labeled above each pair.

With \textbf{full scaffolding} (descriptive + demonstrative cues), the standard model slightly outperforms: Standard=$0.88$, Reasoning=$0.89$, $\Delta=-0.03$. Removing descriptive cues (\textbf{No Descriptions}) equalizes performance at $\approx 0.98$ for both. When only descriptive cues remain (\textbf{No Demonstrations}), reasoning-augmented models gain a small edge: Reasoning=$0.90$, Standard=$0.94$, $\Delta=+0.03$. The advantage of reasoning becomes pronounced in the \textbf{Raw} condition (no cues): Reasoning=$0.97$, Standard=$1.15$, $\Delta=+0.18$, with reasoning-augmented models also showing narrower variance. This pattern reveals that explicit reasoning primarily benefits performance when prompt scaffolding is sparse—when models lack structured clinical definitions and examples, reasoning helps navigate the ambiguity. Conversely, with strong descriptive cues, standard models can match or even exceed reasoning-augmented ones.

\subsection{Mixed-Effects Analysis: Disentangling Model and Task Factors}
\label{ssec:mixed-effects}

\Cref{fig:items-main} summarizes separate task-level, seed-averaged mixed-effects fits from \Cref{ssec:stats}: one fit for reasoning-augmented models ($N{=}92{,}864$ observations; sessions within patients $n{=}541$, patients $n{=}277$, tasks $n{=}11$, models $n{=}16$) and one fit for non-reasoning models ($N{=}52{,}236$ observations; sessions within patients $n{=}541$, patients $n{=}277$, tasks $n{=}11$, models $n{=}9$).

\paragraph{Model characteristics.}
Larger models yield lower error in both strata (log-parameters, $z$-standardized): reasoning-augmented $\hat\beta=-0.638$ ($t=-4.84$), standard $\hat\beta=-0.206$ ($t=-2.49$). Context length shows divergent effects: positive for reasoning-augmented ($\hat\beta=+0.966$, $t=3.51$) but negligible for standard models ($\hat\beta=-0.093$, $t=-1.92$). Mixture-of-Experts architectures correlate with higher error in both strata (reasoning-augmented: $\hat\beta=+0.525$, $t=2.02$; standard: $\hat\beta=+0.308$, $t=1.73$).

\paragraph{Clinical complexity.}
Higher severity predicts higher error both \emph{between} patients (reasoning-augmented: $\hat\beta=+0.119$, $t=21.4$; standard: $\hat\beta=+0.131$, $t=20.1$) and \emph{within} patients (reasoning-augmented: $\hat\beta=+0.137$, $t=33.7$; standard: $\hat\beta=+0.078$, $t=15.3$). Transcript length shows small effects: between-patient verbosity is slightly harder (reasoning-augmented: $\hat\beta=+0.035$, $t=2.46$; standard: $\hat\beta=+0.105$, $t=6.64$).

\paragraph{Reasoning trace patterns.}
For reasoning-augmented models only, reasoning length exhibits a bidirectional pattern: models/sessions with \emph{typically} longer traces show higher error (between-patient/item $\hat\beta=+0.225$, $t=43.5$), but producing \emph{more} reasoning than usual for a given patient–item correlates with lower error (within-session $\hat\beta=-0.929$, $t=-64.9$). This suggests adaptive reasoning—deploying extra computation when needed—is beneficial, while uniform verbosity correlates with harder cases rather than improved accuracy.

\paragraph{Item-level heterogeneity.}
Random-intercept BLUPs (\Cref{fig:items-main}b) reveal a consistent difficulty hierarchy across architectures. Both reasoning-augmented and standard models find \emph{Inability to Feel} (I8; deviations $+0.19$ and $+0.16$, respectively) and \emph{Apparent Sadness} (I1; $+0.08$ and $+0.18$) the hardest items, while \emph{Inner Tension} (I3; $-0.23$), \emph{Reduced Appetite} (I5; $-0.13$ and $-0.17$), and \emph{Suicidal Thoughts} (I10; $-0.12$ and $-0.16$) are consistently easiest. The broad alignment of these rankings across the two strata indicates that item difficulty is driven primarily by construct properties—concrete behavioral items are easier than subjective affective ones—rather than by architectural differences. The elevated difficulty of \emph{Apparent Sadness} is particularly instructive: despite interview questions explicitly asking patients what others comment on about their appearance, this item retains a fundamental observational component—clinical \textsc{MADRS} administration relies heavily on the clinician's direct visual assessment of facial expression, posture, and nonverbal presentation, modalities entirely absent from text transcripts, suggesting that certain depression symptoms would benefit from multimodal integration of visual and acoustic signals. One notable divergence concerns the rescaled total-score task: reasoning-augmented models show the largest positive deviation ($+0.44$), reflecting the difficulty of \textsc{DTS} scoring discussed in \Cref{ssec:total-performance}, whereas standard models show a negative deviation ($-0.11$), suggesting that \textsc{DTS} prediction is comparatively less penalized in the non-reasoning stratum.

\section{Discussion}
Our comprehensive evaluation of $25$ state-of-the-art open-source LLMs on the \textsc{LlaMADRS} benchmark reveals nuanced patterns about when reasoning augmentation benefits clinical assessment. We find that strong open LLMs achieve clinically moderate item-wise accuracy on \textsc{MADRS} from real interviews, using established interpretability bands for error magnitude~\citep{Montgomery1979,Turkoz2021}. Building on this baseline, the \textsc{ItS} procedure—score items, then sum—substantially reduces total MAE for most models, with the majority achieving acceptable error ($<6$); notably, even when reasoning-augmented models "self-aggregate" in their traces, their \textsc{DTS} predictions remain inferior to summing item estimates. However, this benefit does not extend uniformly to all models: smaller models with limited capacity ($\leq 8$B parameters) may remain in the moderate error band even after aggregation, highlighting the importance of adequate model scale for clinical reliability.

The relationship between reasoning augmentation and performance proves context-dependent. Under anchored, schema-constrained prompts that encode clinical descriptive cues, standard variants frequently match or surpass their reasoning-augmented counterparts. Our ablation study clarifies this pattern: explicit reasoning helps primarily when scaffolding is sparse, aligning with recent reports that reasoning gains are task- and prompt-dependent. With full clinical scaffolding (descriptive and demonstrative cues), standard models achieve $\Delta=-0.03$ advantage; without any cues, reasoning-augmented models gain $\Delta=+0.18$ advantage with reduced variance.

A mixed-effects analysis further contextualizes these results: larger parameter counts predict lower error, context-window length has near-zero effect, and—after adjusting for case mix—both MoE architecture and uniform reasoning augmentation correlate with higher error. Critically, longer \emph{within-session} reasoning traces correlate with lower error ($\hat\beta=-0.929$), while models that \emph{typically} produce longer traces show higher error ($\hat\beta=+0.225$). This bidirectional pattern suggests that adaptive reasoning—deploying extra computation selectively—is beneficial, whereas indiscriminate verbosity indicates difficulty rather than capability.

\paragraph{Clinical implications.}
The consistent achievement of acceptable-to-moderate error bands across strong models suggests potential clinical utility, particularly for the \textsc{ItS} approach (item-level scoring followed by summation). The systematic difficulty pattern across items—with subjective/affective items showing higher error than behavioral items—highlights fundamental challenges in text-based assessment of internal states. This pattern persists regardless of reasoning augmentation, suggesting inherent limitations in inferring mood states from dialogue alone.

\paragraph{Methodological insights.}
Our findings challenge assumptions about the uniform superiority of reasoning augmentation. The effectiveness of descriptive cues—clinical anchors and severity definitions—as the highest-value prompt component suggests that for well-structured clinical tasks with clear rubrics, careful prompt engineering can substitute for explicit reasoning. This has practical implications for deployment: standard models with good prompts may be more efficient than reasoning-augmented models that require additional computation without commensurate gains.

\section{Limitations}
Our analysis is text-only; nonverbal cues and prosody—important for affective items—are not modeled. The \textsc{CAMI} dataset represents inpatient psychiatric settings, potentially limiting generalizability to outpatient or community samples. Future work should explore multimodal integration, cross-population validation, and methods for uncertainty quantification appropriate for clinical deployment.

\section{Conclusion}
We introduced \textsc{LlaMADRS}, a benchmark for evaluating LLM performance on clinical depression assessment using real patient--clinician interviews with $5{,}804$ clinician-rated \textsc{MADRS} annotations. Evaluating $25$ models reveals that strong open LLMs achieve clinically acceptable item-level accuracy, while \textsc{ItS} scoring reduces total MAE by $30$--$70$\% compared to \textsc{DTS} prediction, exposing fundamental limits of end-to-end \textsc{DTS} regression.

Reasoning augmentation benefits are context-dependent: standard models often match reasoning-augmented variants under well-structured prompts ($\Delta=-0.03$), but reasoning provides clear gains with sparse scaffolding ($\Delta=+0.18$). Mixed-effects analyses identify model scale and reasoning length as key predictors, with reasoning tokens showing the largest effect ($\beta=-0.929$). These findings challenge assumptions about reasoning augmentation and establish a foundation for automated yet interpretable clinical assessment.

\section*{Acknowledgments}
This material is based upon work partially supported by National Institutes of Health awards R01MH125740, R01MH132225, U01MH136535, and R21MH130767. Any opinions, findings, conclusions, or recommendations expressed in this material are those of the authors and do not necessarily reflect the views of the sponsors, and no official endorsement should be inferred.
\clearpage
\bibliography{main}
\bibliographystyle{acl_natbib}

\iftaclpubformat

\onecolumn
\appendix

\section{Prompt Templates}
\label{sec:appendix-prompts}

\definecolor{descCol}{HTML}{1A6B8A}   
\definecolor{descBg}{HTML}{EBF5FA}    
\definecolor{demoCol}{HTML}{B8600A}   
\definecolor{demoBg}{HTML}{FDF3E7}    
\definecolor{schemCol}{HTML}{555555}  

Each prompt follows the three-component structure from \S\ref{ssec:prompt-framework}: (1)~\textcolor{descCol}{descriptive cues}, (2)~\textcolor{demoCol}{demonstrative cues}, and (3)~JSON output schema. Ablation configurations (\S\ref{ssec:ablation-methods}) selectively remove (1) and/or (2). Full transcripts are provided in the actual prompts; excerpts below are abbreviated for space.

\subsection{\textsc{DTS} Prompt}
\label{ssec:total-prompt}

\begin{tcolorbox}[colback=gray!3, colframe=gray!50, coltitle=black, title={\textbf{\textsc{DTS}: Total Score (0--60)}}, fonttitle=\small\bfseries, breakable, boxsep=3pt, left=4pt, right=4pt, top=3pt, bottom=3pt]
\scriptsize\ttfamily
Task: Analyze a diarized transcript of a psychiatric session where the Montgomery-\AA{}sberg Depression Rating Scale (MADRS) questionnaire is being administered. Predict the total MADRS score (0--60) that the practitioner would likely give based on the patient's responses and the conversation. Remember that the rating is for the last week, not based on the patient's history or general condition.
\begin{tcolorbox}[colback=descBg, colframe=descBg, boxsep=2pt, left=3pt, right=3pt, bottom=2pt, sharp corners, boxrule=0pt]
\textcolor{descCol}{\textbf{[Descriptive Cues]}}\\[3pt]
The MADRS consists of 10 items, each scored from 0--6:\\
1.~Apparent Sadness, 2.~Reported Sadness, 3.~Inner Tension, 4.~Reduced Sleep, 5.~Reduced Appetite, 6.~Concentration Difficulties, 7.~Lassitude, 8.~Inability to Feel, 9.~Pessimistic Thoughts, 10.~Suicidal Thoughts\\[3pt]
Severity ranges: 0--6 Normal; 7--19 Mild; 20--34 Moderate; 35--60 Severe.
\end{tcolorbox}
Instructions:\\
1. Read the provided interview transcript.\\
2. Predict the total MADRS score (0--60) based on the overall conversation.\\
3. Provide a brief explanation (2--3 sentences) for your predicted score.\\[3pt]
Output Schema:
\{rating: 0--60, explanation: 2--3 sentences\}
\begin{tcolorbox}[colback=demoBg, colframe=demoBg, boxsep=2pt, left=3pt, right=3pt, bottom=2pt, sharp corners, boxrule=0pt]
\scriptsize\ttfamily
\textcolor{demoCol}{\textbf{[Demonstrative Cues]}} --- 3 examples at different severity levels\\[2pt]
{\normalfont\itshape Example 1 (No Depression):}\\
Transcript: \texttt{<full transcript>}\\
Rating: 4\\
Explanation: [2--3 sentences]\\[3pt]
{\normalfont\itshape Example 2 (Moderate):}\\
Transcript: \texttt{<full transcript>}\\
Rating: 25\\
Explanation: [2--3 sentences]\\[3pt]
{\normalfont\itshape Example 3 (Severe):}\\
Transcript: \texttt{<full transcript>}\\
Rating: 51\\
Explanation: [2--3 sentences]
\end{tcolorbox}
\end{tcolorbox}

\subsection{Item-Level Prediction Prompt}
\label{ssec:item-prompt}

Each of the ten \textsc{MADRS} items uses the same template with item-specific definitions, questions, anchors, and seven demonstrative examples (one per anchor 0--6). Below we illustrate with \emph{Reported Sadness} (Item~2).

\begin{tcolorbox}[colback=gray!3, colframe=gray!50, coltitle=black, title={\textbf{Item Prompt: Reported Sadness (0--6)}}, fonttitle=\small\bfseries, breakable, boxsep=3pt, left=4pt, right=4pt, top=3pt, bottom=3pt]
\scriptsize\ttfamily
Task: Analyze a diarized transcript of a psychiatric session where the MADRS is being administered. Predict the rating (0--6) that the practitioner would likely give for the specified MADRS item based on the patient's responses and the conversation. Remember that the rating is for the last week, not based on the patient's history or general condition. Focus on the questions that are directly related to the MADRS item and the patient's answers.\\[3pt]
MADRS Item: Reported Sadness
\begin{tcolorbox}[colback=descBg, colframe=descBg, boxsep=2pt, left=3pt, right=3pt, bottom=2pt, sharp corners, boxrule=0pt]
\textcolor{descCol}{\textbf{[Descriptive Cues]}}\\[3pt]
\scriptsize\ttfamily
Description: Representing reports of depressed mood, regardless of whether it is reflected in appearance or not. Includes low spirits, despondency or the feeling of being beyond help and without hope. Rate according to intensity, duration and the extent to which the mood is reported to be influenced by events.\\[3pt]
Rating Scale:\\
0 -- Occasional sadness in keeping with the circumstances.\\
1 -- Between 0 and 2.\\
2 -- Sad or low but brightens up without difficulty.\\
3 -- Between 2 and 4.\\
4 -- Pervasive feelings of sadness or gloominess. The mood is still influenced by external circumstances.\\
5 -- Between 4 and 6.\\
6 -- Continuous or unvarying sadness, misery or despondency.
\end{tcolorbox}
\vspace{3pt}
Instructions:\\
1. Read the provided interview transcript.\\
2. Rate the interviewee's reported sadness on a scale of 0--6.\\
3. Provide a brief explanation (2--3 sentences) for your rating.\\
4. List 2--3 key utterances from the conversation supporting your assessment.\\[3pt]
Output Schema:
\{rating: 0--6, explanation: 2--3 sentences, key\_utterances: [line numbers], most\_relevant\_question: [...]\}

\begin{tcolorbox}[colback=demoBg, colframe=demoBg, boxsep=2pt, left=3pt, right=3pt, bottom=2pt, sharp corners, boxrule=0pt]
\textcolor{demoCol}{\textbf{[Demonstrative Cues]}} --- 7 examples (0--6)\\[2pt]
\scriptsize\ttfamily
Transcript: \texttt{<item-segmented transcript>}\\
Rating: 0\\
Explanation: [2--3 sentences]\\
Key Utterances: [line nos.]\\
Most Relevant Question: [...]\\[3pt]

...\\[3pt]

Transcript: \texttt{<item-segmented transcript>}\\
Rating: 6\\ 
Explanation: [2--3 sentences]\\
Key Utterances: [line nos.]\\ 
Most Relevant Question: [...]
\end{tcolorbox}
\end{tcolorbox}

\end{document}